\documentclass[acmtog, authorversion]{acmart}

\usepackage{booktabs} % For formal tables

\setcitestyle{square}

\usepackage{graphicx}
\usepackage{amsmath,bm,bbm,dsfont}
\usepackage{subfigure}
\usepackage{caption}
\setcopyright{usgovmixed}

\newenvironment{myitemize}[1][]{%%%%%
\begin{list}{{#1}} %%
    {
     \setlength{\leftmargin}{0cm}      %
     \setlength{\parsep}{0ex}          %
     \setlength{\topsep}{0bp}          %
     \setlength{\itemsep}{0ex}         %
     \setlength{\labelsep}{0.2em}      %
     \setlength{\itemindent}{2em}      %
     \setlength{\listparindent}{1.5em} %
    }}
{\end{list}}%%%%%
\begin{document}
% Title portion
\title{Normal Image Manipulation for Bas-relief Generation with Hybrid Styles}

\author{Zhongping Ji}
\orcid{1234-5678-9012-3456}
\affiliation{%
  \institution{Hangzhou Dianzi University}
  \city{Hangzhou}
  \country{China}}
\author{Xianfang Sun}
\affiliation{%
  \institution{Cardiff University}
  \city{Cardiff}
  \country{United Kingdom}
}
\author{Weiyin Ma}
\affiliation{%
 \institution{City University of Hong Kong}
 \city{Hong Kong}
 \country{China}
 }

\renewcommand\shortauthors{Ji, Z.P. et al}

\begin{abstract}
We introduce a normal-based bas-relief generation and stylization method
which is motivated by the recent advancement in this topic.
Creating bas-relief from normal images has successfully
facilitated bas-relief modeling in image space.
However, the use of normal images in previous work is often restricted to certain type of operations only.
This paper is intended to extend normal-based methods and construct bas-reliefs from normal images in a versatile way.
Our method can not only generate a new normal image by combining various frequencies of existing normal images and details transferring,
but also build bas-reliefs from a single RGB image and its edge-based sketch image.
In addition, we introduce an auxiliary function to represent
a smooth base surface and generate a layered global shape.
To integrate above considerations into our framework,
we formulate the bas-relief generation as a variational problem
which can be solved by a screened Poisson equation.
Some advantages of our method are that it expands the bas-relief shape space and generates diversified styles of results,
and that it is capable of transferring details from one region to other regions.
Our method is easy to implement, and produces good-quality bas-relief models.
We experiment our method on a range of normal images
and it compares favorably to other popular classic and state-of-the-art methods.
\end{abstract}



\maketitle

\section{Introduction}

Relief, commonly used for thousands of years,
is a form of sculpture in which a solid piece of material is carved
so that figures slightly emerge from a background.
Bas-relief is a type of relief (sculpture) that has less depth to the faces and figures than they actually have, when measured proportionately (to scale).
Even with the development of computer-aided-design, the design of bas-reliefs remains mainly in the hands of artists.
Recently, the problem of automatic generation of bas-reliefs from 3D input scenes has received great attention.
To simplify the problem and to borrow some approaches developed for image processing,
bas-reliefs generated using previous methods usually are represented as height fields which have a single $z$ depth for each $(x,y)$ position.
Consequently, the key ingredient for the converting 3D scenes to bas-reliefs
is the compression of the height field sampled from the input 3D scenes.
Most of previous methods focused on designing sophisticated non-linear depth compression algorithms
to fulfill the task, while some other methods focused on reducing the computation cost.
Recently, Ji et al. presented a novel method for bas-relief modeling \cite{JMS14}.
Unlike previous work, their method designed bas-reliefs in normal image space instead of in object space.
Given a composite normal image, the problem involves generating
a discontinuity-free depth field with high compression of depth data
while preserving fine details.
However, their work attempted to generate bas-reliefs
with the original normals sampled from an orthogonal view.
This paper further extends their method in two different perspectives,
with added capability of hierarchical editing of the normal image and
the introduction of an auxiliary function  for representing the global smooth base surface.
Given a normal image, we decompose it into different layers which may be edited
and be composed again if necessary.
Our method then attempts to construct bas-reliefs from the resulting normal image
which has different levels of details.
In addition, previous work seldom assign a global shape at this stage to control the basic shape of the bas-reliefs.
Our work makes an attempt to advance forward in this direction.
\noindent \textbf{\\Contributions.}
This paper develops a simple but effective method to solve some problems existing in the recent work \cite{JMS14}.
To utilize normals more reasonably,
we develop a $DoG$-like filter to decompose a given normal image.
In addition, a variational formulation with a data fidelity term and a regularization term is proposed
to control the global appearance of the resulting bas-relief.
The main contributions of this paper are summarized as follows:
\begin{myitemize}[\textbullet]

\item \textbf{Normal Decomposition}.
We propose a $DoG$-like filter to decompose normal image to its high-frequency and low-frequency components.
Once high-frequency and low-frequency components are decomposed,
one can transplant the high-frequency components to other normal images in a way similar to geometric processing.

\item \textbf{Normal Composition}.

Our method blends two normal fields in a more reasonable way,
that is in normal space instead of in image space.
Consequently, details are added into a base layer by taking the orientations of normals
into account.

\item \textbf{Hybrid Stylization}.
We also introduce an auxiliary function $h(u,v)$ to construct a base surface
to convey different 3D impressions of the resulting bas-reliefs.
The base surface includes smooth shape and step-shaped surface,
which makes our method a more effective and flexible bas-relief modeling tool.

\item \textbf{Bas-relief from Images}.
We also propose a two-scale algorithm for producing a relief shape from a single general image.
Our algorithm calculates a detail normal image from the input image,
constructs a base normal image starting from a few strokes,
and finally combines them in the normal domain to produce a plausible bas-relief.

\end{myitemize}

This paper is organized as follows:
Section 2 describes related work and summarizes state-of-the-art approaches.
In Section 3 we give an overview of our algorithm.
In Section 4 we introduce a $DoG$-like filter to extract higher-frequency details.
In Section 5, we transform a single image to a plausible bas-relief.
We formulate the problem using a variational framework in Section 6.
Results and comparisons are shown in Section 7.
We conclude the paper with future work in Section 8.

%-------------------------------------------------------------------------

\section{Related Work}
The design of bas-reliefs has been an interesting topic in computer graphics in the past two decades.
In this section, we provide a review of bas-relief generation methods that are most relevant to ours.

%1, 3D
Early work on bas-relief modeling mainly focused on generating bas-relief models from 3D models or scenes.
Cignoni et al. treat the bas-relief generation
as a problem of compressing the depth of a 3D scene onto a viewing plane \cite{CMS97}.
Their principle rule is to treat a 3D scene as a height field from the point of view of the
camera, which is followed by the subsequent literature.
The majority of previous methods encode the height field as an image.
Following this approach, the research into digital bas-relief modeling
is largely inspired by corresponding work on 2D images,
such as some approaches developed for tone mapping of
high dynamic range (HDR) images and histogram equalization.
For bas-reliefs, depths from a 3D scene are represented by the intensities in a HDR image.
Weyrich et al. propose an HDR-based approach
for constructing digital bas-reliefs from 3D scenes \cite{WDBRF07}.
Their work truncates and compresses the gradient magnitude to remove depth discontinuities
in a similar way as the work on 2D images \cite{FLW02}.
Kerber et al. propose a feature preserving bas-relief generation approach
combined with linear rescaling and unsharp masking of gradient magnitude \cite{kbs07}.
An improvement on this approach is proposed in \cite{Ker07},
which rescales the gradient nonlinearly.
Using four parameters, one can steer the compression ratio and
the amount of details expected in the output.
Kerber et al. also present a filtering approach
which preserves curvature extrema during the compression process \cite{KTRAH09}.
In this way, it is possible to handle complex scenes with fine geometric details.
Song et al. create bas-reliefs on the discrete differential coordinate domain,
combining the mesh saliency and the shape exaggeration \cite{SBS07}.
Inspired by the relations among HDR, histogram equalization and bas-relief generation,
Sun et al. apply an adaptive histogram equalization method to the depth compression,
which provides a new algorithm on bas-relief generation \cite{SPRF09}.
This approach produces high quality bas-relief and preserves surface features well.
Bian and Hu propose an approach based on gradient
compression and Laplacian sharpening, which produces bas-reliefs with well-preserved details \cite{BianH11}.

% 2, image
Other than 3D scenes, some work has been reported on generating bas-reliefs from 2D images.
Li et al. present a special approach for bas-relief estimation
from a single special image which is called a rubbing image \cite{LWYM12}.
They aim at restoring brick-and-stone bas-reliefs from their rubbing images in a visually plausible manner.
Wu et al. develop an approach for producing bas-reliefs from human face images \cite{wmr13}.
They first create a bas-relief image from a human face image,
and then use a shape-from-shading (SfS) approach on the bas-relief image
to construct a corresponding bas-relief.
They train and use a neural network to map a human face image to a bas-relief image,
and apply image relighting technique to generate relit human face images
for bas-relief reconstruction.
Zhang et al. propose a mesh-based modeling system to generate Chinese calligraphy reliefs from a single image \cite{ZCLJZ18}.
The relief is constructed by combining a homogeneous height field and an inhomogeneous height field together via a nonlinear compression function.
However, these methods are often restricted to generating special type of bas-reliefs from special images.
% 3, GPU
Another type of methods \cite{KTA2010,zzz13,JSLW14} focus on reducing the computation cost
to present real-time interactive tools designed to support artists and interested designers
in creating bas-relief using 3D scenes.
In addition, the work \cite{JSLW14} is also suitable to create bas-reliefs with different styles.

Recently, several new techniques are developed in this topic.
S\'{y}kora et al. present an interactive approach for generating bas-relief sculptures
with global illumination rendering of hand-drawn characters using a set of annotations \cite{Sykora14}.
Sch\"uller et al. propose a generalization of bas-reliefs,
which is capable of creating surfaces that depict certain shapes from prescribed viewpoints \cite{SPS14}.
Their method can be applied to generate standard bas-reliefs,
optical illusions and carving of complex geometries.
Zhang et al. propose a bas-relief generation method through gradient-based mesh deformation \cite{ZZLLZ15}. Unlike image-based methods, their method works in object space and
retains the mesh topology during geometric processing.
Through gradient manipulation of the input meshes,
their method is capable of constructing bas-reliefs on planar surfaces and on curved surfaces directly.
Zhang et al. propose a novel approach for producing bas-reliefs with desired appearance under illumination \cite{ZZWC2016}.
Ji et al. develop a novel modeling technique for creating digital bas-reliefs in normal domain \cite{JMS14}.
The most import feature of their approach is that it is capable of producing
different styles of bas-reliefs and permits the design of
bas-reliefs in normal image space rather than in object space.
In this paper, we further enrich this research topic
in creating bas-reliefs in normal image space.
Although the bilateral filter has also been used in their work, it is just used to smooth
the normal image for the application of bas-reliefs modeling from general images.
In this paper, the bilateral filter will be used to define a decomposition operation which extracts information of various frequencies from a normal image.
Furthermore, we develop a layer-based editing approach for normal images and integrate an auxiliary function
to control the global appearance of the resulting bas-relief.
Due to the normal decomposition-and-composition operations and the auxiliary function,
our method is capable of producing more styles of bas-reliefs.

\section{Problems and Motivation}
Ji et al. develop a normal-based modeling technique for creating digital bas-reliefs \cite{JMS14}.
The most important feature of their approach is that it permits the design of bas-reliefs in normal image space rather than in object space.
We also design bas-reliefs in normal image space.
However, we propose significantly different techniques to manipulate normal images, aiming at solving the following problems on connection with their approach.
\begin{myitemize}[\textbullet]

\item In their layer-based framework, one can reuse existing normal images to design new bas-reliefs by a cut-and-paste operation.
However, the uppermost layer will override other ones \textbf{which implies that it is not suitable for detail transferring between layers}.

\item A threshold $\theta$ is introduced to control the resulting height field,
which produces different visual styles of bas-reliefs.
New techniques are needed to enrich the visual complexity and the range of shapes of resulting bas-reliefs.

\item In their bas-relief modeling implementation, a general colored image can be used in the process of the bas-relief generation.
Their method focuses on converting general images into geometry textures,
but lacks a mechanism of building a plausible bas-relief from a single colored image.

\end{myitemize}

One important contribution of this work is to propose a decomposition filter for normal images.
The purpose of the introduction of this filter is twofold.
On one hand, we want to extract information of various frequencies from a given normal image.
The original normal image encodes full frequencies of depth information of an underlying 3D scene.
We attempt to simulate various effects which correspond to the information of different frequencies.
On the other hand, we also want to develop a tool to edit a normal image in a two-scale manner.
Once a normal image is decomposed into two layers,
 one can edit either or both layers and combine them again to obtain a new normal image.
Furthermore, one can also transfer high-frequency patches extracted from a normal image to
other normal images.
In short, one of our purposes is to construct various bas-reliefs from given normals in a flexible way.

Another contribution of this work is to propose a two-scale approach for creating bas-reliefs from a single general image.
The key idea is to build both a base surface representing the general smooth shape and feature details in the normal domain.
Our algorithm calculates a detail normal image from a given image directly,
constructs a base normal image using a few interactive operations,
and finally combines them using the proposed composition operation to produce a plausible bas-relief.

In addition, to control the global appearance of the resulting bas-relief,
a variational formulation with a data fidelity term and a regularization term is proposed in this paper.
Specifically, we introduce an auxiliary function to assign a smooth base shape or a layered global shape.

The steps of our algorithm are listed as follows,
\begin{itemize}
\item to decompose a normal image into two layers, including a high-frequency component and a low-frequency one, and to construct two normal image layers from a single general image;
\item to edit either or both layers and combine them with transferred details by the composition operation if necessary;
\item to use a smooth or step function to indicate a global shape if necessary;
\item to generate a bas-relief by solving a screened Poisson equation.
\end{itemize}
Some of the above steps are optional and details of these steps are described in following sections.

\section{Normal Image Operations}

\subsection{Decomposition-and-Composition of Normals}

An important step of our approach is to decompose a normal image into two layers,
including a detail image and a base image.
To this end, we define a decomposition filter based on a bilateral filter.
Details from the normal image are extracted through the difference of the original normal image and a smoothed one.
First, following the definition of a bilateral filter for images \cite{TM98},
our bilateral filter for a pixel $p$ in a normal image $N$ is defined as:
\[
N_{\Sigma}( p )  = \frac{1}{{k_p }}\sum\limits_{q \in N}
{N(q) w_c(||p - q||)w_s(||N(p)-N(q) ||)},
\]
where $\Sigma = (\sigma_c,\sigma_s)$, $w_c(x) = exp(-x^2/2{\sigma_c}^2)$ is the closeness function,
$w_s(x) = exp(-x^2/2{\sigma_s}^2)$ is the similarity function,
and $k_p$ is a scaling factor that normalizes $N_\Sigma(p)$ to a unit vector.

Unlike general RGB images, a pixel in a normal image indicates a normal vector.
Thus, when we consider the difference between two normal vectors,
we should take the orientation into account.
Specifically, the rotation, between two normal vectors from the base layer and the detail layer respectively, is used to define the difference of two vectors.
We first record the rotation from a normal vector $n$ in base layer to vector $z=(0,0,1)$
along the axis $n \times z$,
and then use the rotation to deform the corresponding normal vector in detail layer along the same axis.
Specifically, we define the normal decomposition filter as follows,

\begin{equation}{\label{eqn:nfilter}}
\begin{array}{rcl}
D_{\sigma_c,\sigma_s}(p)
= N(p)
\; \ominus \; N_{\sigma_{c},\sigma_{s}}(p),
\end{array}
\end{equation}
where 
the operator $\ominus$ is not a general minus operation $-$ of two vectors.
the operator $\ominus$ measures the deviation between two normal vectors,
and it is defined as follows,
\begin{equation}{\label{eqn:ominus}}
\begin{array}{rcl}
n_1 \ominus n_2
=Q(n_2,n_0) N_1 Q(n_2,n_0)^{-1},
\end{array}
\end{equation}
where $n_0$ is a constant vector $[0, 0, 1]$, $N_1=[0,n_1]$,
and $Q(x,y)$ is a quaternion representing the rotation from vector $x$ to vector $y$
along the axis $x \times y$.

Similar to the $DoG$ filter, our filter is an approximate band-pass operator
which is used for revealing the normal difference through a certain band.
The $N(p)$ in Equation (\ref{eqn:nfilter}) indicates the original normal.
If necessary, one can set a proper $\sigma = (\sigma_{c'},\sigma_{s'})$
to eliminate noises or outliers existing in the original normal image,
that is $N(p) = N_{\sigma_{c'},\sigma_{s'}}(p)$.

\begin{figure}
  \centering
  \subfigure[]{
    \label{fig:budda1:a} 
    \includegraphics[angle=0,width=0.80in]{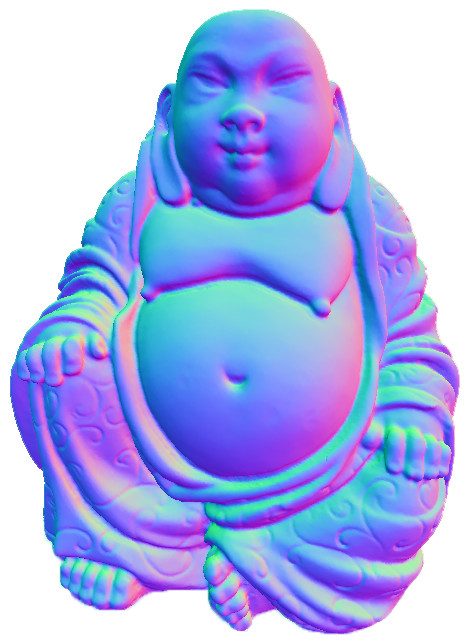}}
  \hspace{-0.1in}
  \subfigure[]{
    \label{fig:budda1:b} 
    \includegraphics[angle=0,width=0.80in]{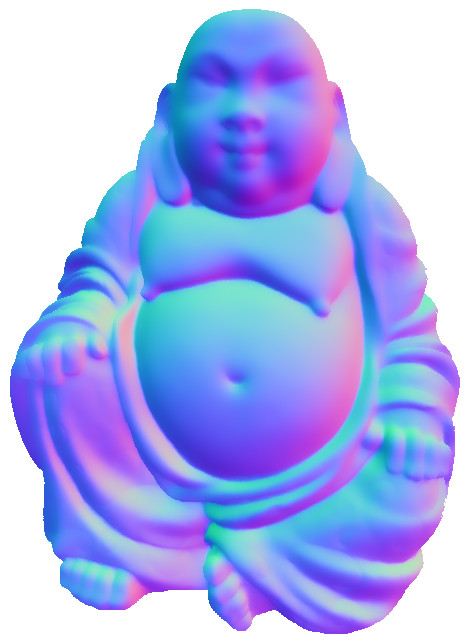}}
    \hspace{-0.1in}
  \subfigure[]{
    \label{fig:budda1:c} 
    \includegraphics[angle=0,width=0.80in]{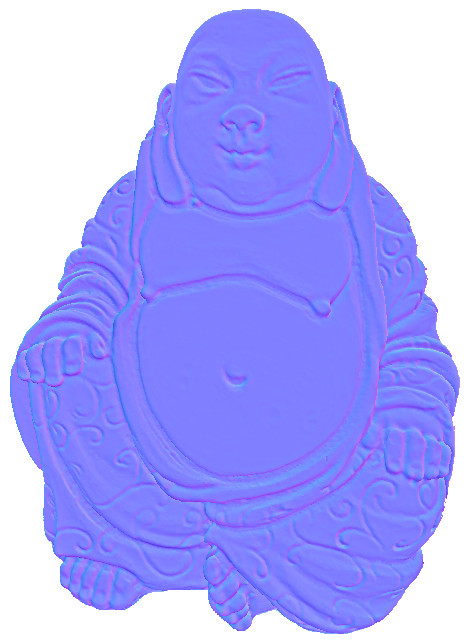}}
    \hspace{-0.1in}
  \subfigure[]{
    \label{fig:budda1:d} 
    \includegraphics[angle=0,width=0.80in]{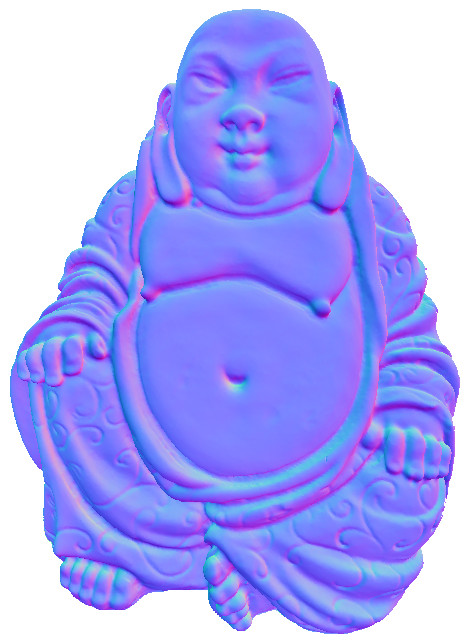}}
  \caption{\label{fig:budda1}
Illustration of our decomposition filters with different band-pass widths:
  (a) an input normal image;
  (b) the smoothed version with parameters $\sigma_c = 3, \sigma_s = 0.9$;
  (c) extracted details by the difference of (a) and (b);
  and (d) extracted details using parameters $\sigma_c = 15, \sigma_s = 0.9$.
}
\end{figure}

\begin{figure}
  \centering
  \subfigure[]{
    \label{fig:details:a} 
    \includegraphics[angle=0,width=0.80in]{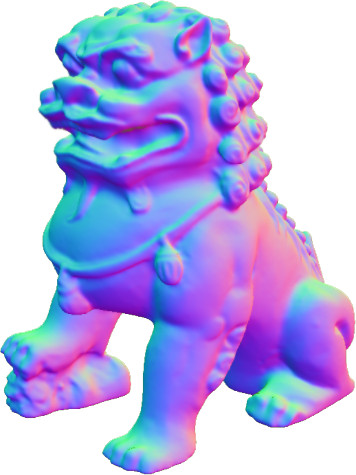}}
  \hspace{-0.0782in}
  \subfigure[]{
    \label{fig:details:b} 
    \includegraphics[angle=0,width=0.80in]{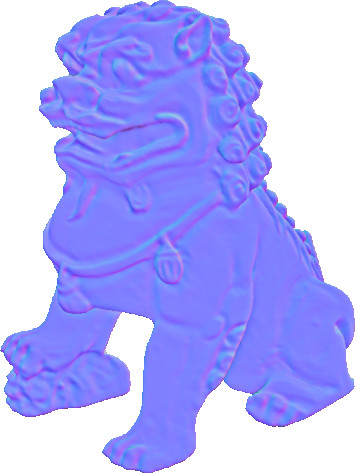}}
    \hspace{-0.0782in}
  \subfigure[]{
    \label{fig:details:c} 
    \includegraphics[angle=0,width=0.80in]{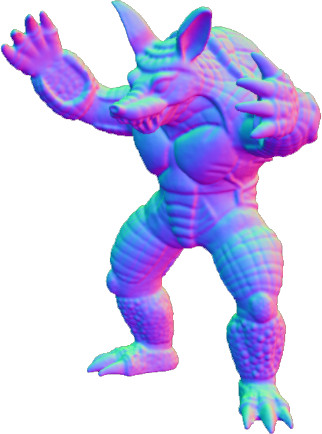}}
    \hspace{-0.0782in}
  \subfigure[]{
    \label{fig:details:d} 
    \includegraphics[angle=0,width=0.80in]{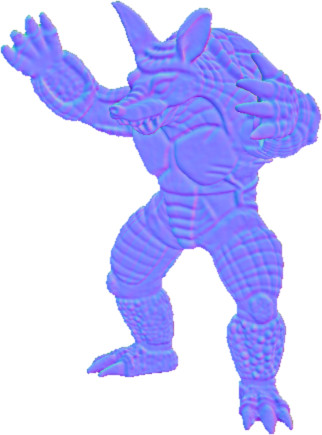}}
  \caption{\label{fig:details}
Details extracted from two normal images using our decomposition filters.
  (a) An input normal image;
  (b) $\sigma_c=3,\sigma_s=0.9$;
  (c) an input normal image;
  and (d) $\sigma_c=3,\sigma_s=0.9$.
}
\end{figure}

To reduce the number of user-specified parameters,
we fix the parameter $\sigma_s = 0.9$ which works well in our experiments.
The output of the decomposition filter includes a detail normal image and a base normal image.
An illustration of our band-pass filter with different band-pass widths are shown in Figure \ref{fig:budda1}.
As can be seen in the figures,
the image smoothed by the bilateral filter (see Figure \ref{fig:budda1:b}) discarded the small-scale fine details while preserved the large-scale shape.
In addition, the output detail normal image varies with different band-pass widths as can be seen in Figure \ref{fig:budda1:c} and Figure \ref{fig:budda1:d}.
Two other examples with the same band-pass width are shown in Figure \ref{fig:details}.

%%% Figure
%%%

%%% Figure
%%%

Once the detail and base normal images are given after the normal decomposition step,
one can further process them separately.
The processed images could be further combined into a new normal image again if necessary.
In this paper, we combine two images with consideration of the orientation of involved normals.
Specifically, we define the composition operation as the inverse operation of the above normal decomposition,
\begin{equation}{\label{eqn:ncomb}}
\begin{array}{rcl}
N_c(p)
= N_d(p)\; \oplus \; N_b(p),
\end{array}
\end{equation}
where $N_d(p)$ and $N_b(p)$ denotes the detail normal and base normal respectively, and $\oplus$ is defined as follows,
\begin{equation}{\label{eqn:oplus}}
\begin{array}{rcl}
n_1 \oplus n_2
=Q(n_0,n_2) N_1 Q(n_0,n_2)^{-1},
\end{array}
\end{equation}
where $n_0$, $N_1$ and $Q(x,y)$ are the same as mentioned above.

\subsection{Hybrid Stylization}
\begin{figure}
  \centering
  \subfigure[]{
    \label{fig:mixture:a} 
    \includegraphics[angle=0,width=0.68in]{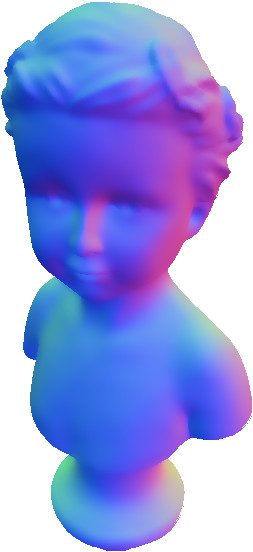}}
    \hspace{0.08in}
  \subfigure[]{
    \label{fig:mixture:b} 
    \includegraphics[angle=0,width=0.68in]{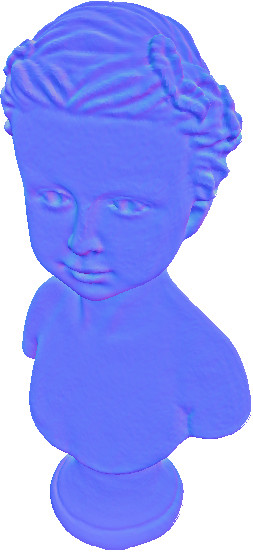}}
    \hspace{0.08in}
  \subfigure[]{
    \label{fig:mixture:c} 
    \includegraphics[angle=0,width=0.68in]{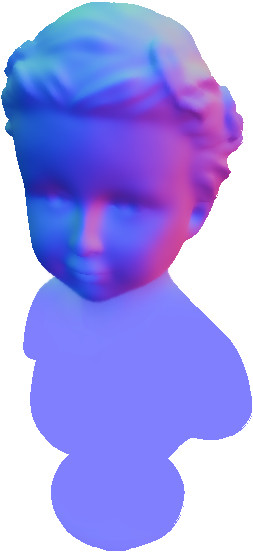}}
    \hspace{0.08in}
  \subfigure[]{
    \label{fig:mixture:d} 
    \includegraphics[angle=0,width=0.68in]{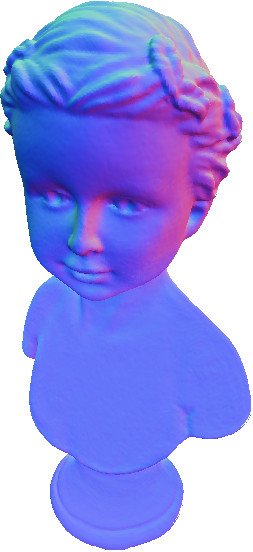}}
  \caption{\label{fig:mixture}
Illustration of combining a base normal image and a detail normal image:
  (a) and (b) a base image and a detail image produced by our decomposition filter;
  (c) a new base normal image by smoothing (a) partially;
  and (d) the mixed result of (b) and (c) generated by our normal composition operation.
}
\end{figure}

To broaden the range of visual effects, our method further blends the high-frequency and low-frequency components in a more flexible way.
Especially, one can suppress the low-frequency layer partially or enhance the high-frequency layer if necessary and then blend them again.
By using these kinds of operations,
our method can easily produce hybrid style of bas-reliefs.
Although the work \cite{JMS14} can also create a similar style,
but their method needs to define an index set using a mask image and compress the heights of different parts carefully.
However, our method can handle the different frequencies directly,
without any mask images and post-processing.
An example is shown in Figure \ref{fig:mixture}.
We edit the base normal image partially, and then combine it with a detail normal image
to form a flattened body and a prominent head.
It finally yields a bas-relief with a special visual effect which is shown in Figure \ref{fig:examples:i}.

%%% Figure
%%%

\begin{figure}
  \centering
  \subfigure[]{
    \label{fig:mixture2:a} 
    \includegraphics[angle=0,width=1.08in]{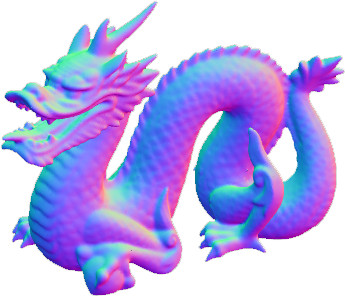}}
  \hspace{-0.08in}
  \subfigure[]{
    \label{fig:mixture2:b} 
    \includegraphics[angle=0,width=1.08in]{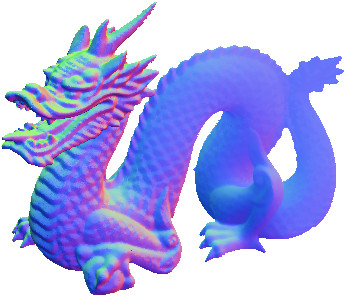}}
    \hspace{-0.08in}
  \subfigure[]{
    \label{fig:mixture2:c} 
    \includegraphics[angle=0,width=1.08in]{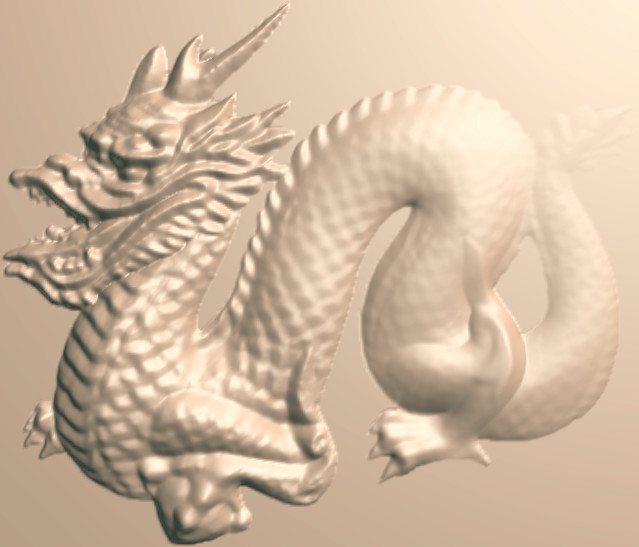}}
  \caption{\label{fig:mixture2}
Illustration of another hybrid bas-relief example produced by feature boosting and layers blending:
  (a) a normal image sampled from a dragon model;
  (b) a revised normal image by a series of operations (decomposing, editing, feature boosting and layers blending);
  and (c) the resulting bas-relief.
}
\end{figure}

Once the normal images are extracted through the decomposition filter,
we can further apply a non-linear tuning function $R(\theta)$ to the rotation angle if necessary.
We found that
\begin{equation}{\label{eqn:rescale}}
R(\theta) = \beta \theta_{m} (\theta/\theta_{m})^{\gamma}
\end{equation}
attenuates or boosts the detailed features in a simple and effective way.
The parameter $\theta$ denotes the rotation angle from the vector $z=[0,0,1]$ to a normal vector $n$ in detail normal image along the axis $z \times n$,
$\theta_{m}$ denotes the maximum of $\theta$.
$\beta$ and $\gamma$ are usually set to values between $0$ and $2$ in our experiments.
Figure \ref{fig:mixture2} shows an example produced by applying the tuning function to a detail normal image
extracted from a dragon model (see Figure \ref{fig:mixture2:a}).
%%% Figure
%%%

\subsection{Detail Transferring}

By means of our composition operation, one can easily transfer detail patches into another place in  a reasonable way.
In doing so, we treat the detail patch as a detail image, and the normal image as base image.

\begin{figure}
  \centering
  \subfigure[]{
    \label{fig:buddabelly:a} 
    \includegraphics[angle=0,width=0.80in]{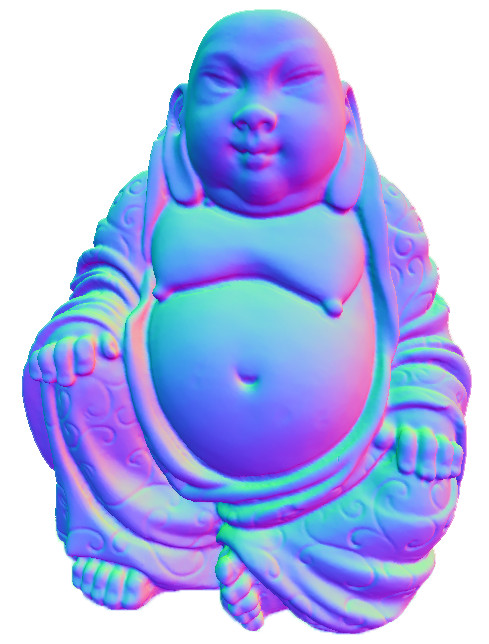}}
  \hspace{-0.0782in}
  \subfigure[]{
    \label{fig:buddabelly:b} 
    \includegraphics[angle=0,width=0.80in]{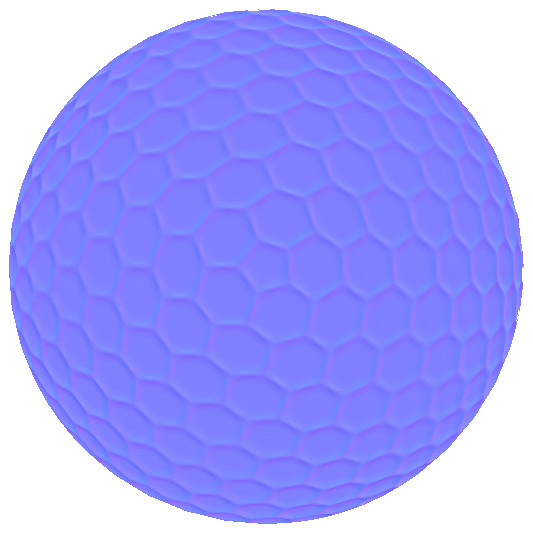}}
    \hspace{-0.0782in}
  \subfigure[]{
    \label{fig:buddabelly:c} 
    \includegraphics[angle=0,width=0.80in]{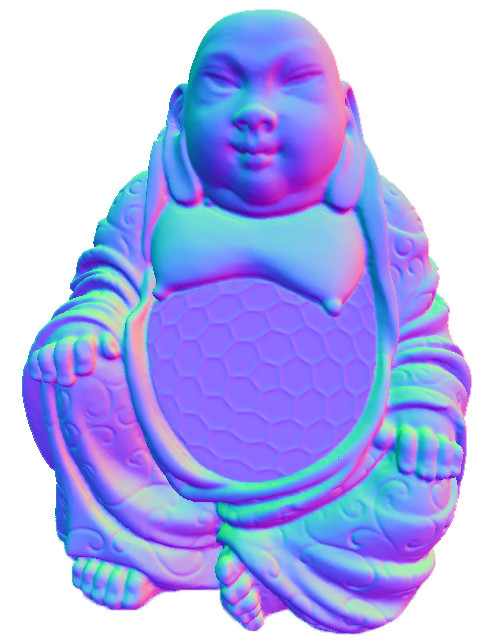}}
    \hspace{-0.0782in}
  \subfigure[]{
    \label{fig:buddabelly:d} 
    \includegraphics[angle=0,width=0.80in]{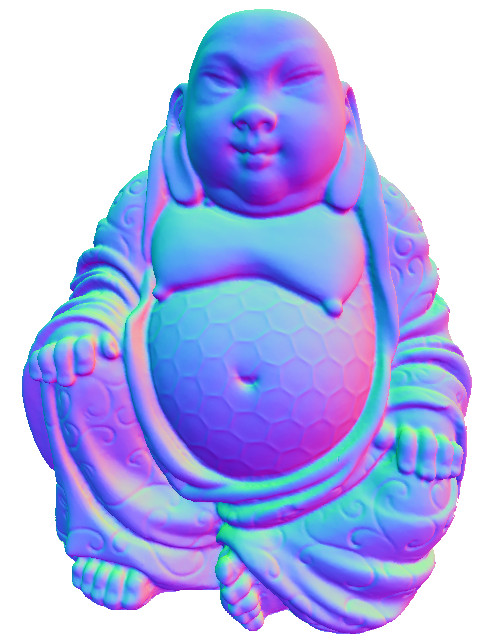}}
  \caption{\label{fig:buddabelly}
Illustration of transferring a detail patch into another normal image:
  (a) a normal image;
  (b) a detail normal image extracted from a golf model;
  (c) a patch of (b) in direct overlapping with (a);
  and (d) the resulting normal image after transferring the patch details to (a).
}
\end{figure}

Figure \ref{fig:buddabelly} gives an example to demonstrate the ability to make details grow on a base surface.
Figure \ref{fig:buddabelly:c} indicates that we let the detail patch directly overlay the base normal image which is equivalent to the cut-and-paste operation in \cite{JMS14}.
Our transferring result is shown in Figure \ref{fig:buddabelly:d}.
As seen from Figure \ref{fig:buddabelly}, the front patch grows well along the belly of the Buddha model.
The resulting bas-relief model is shown in Figure \ref{fig:examples:g}.

\begin{figure}
  \centering
  \subfigure[]{
    \label{fig:hand_dragon:a} 
    \includegraphics[angle=0,width=0.820in]{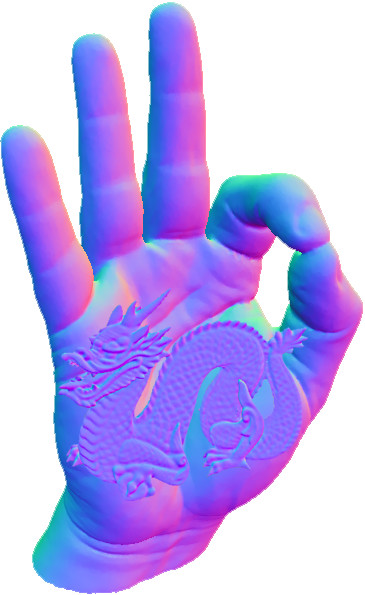}}
  \hspace{-0.12in}
  \subfigure[]{
    \label{fig:hand_dragon:b} 
    \includegraphics[angle=0,width=0.820in]{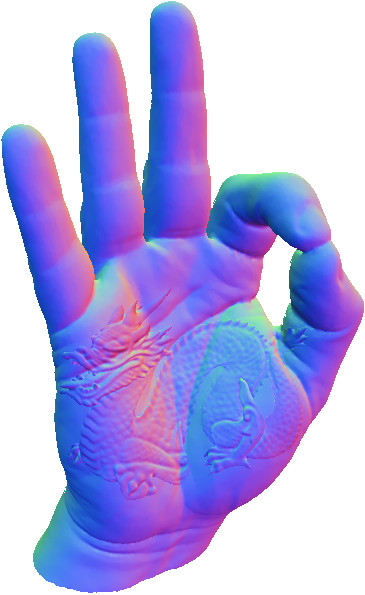}}
    \hspace{-0.12in}
  \subfigure[]{
    \label{fig:hand_dragon:c} 
    \includegraphics[angle=0,width=0.8220in]{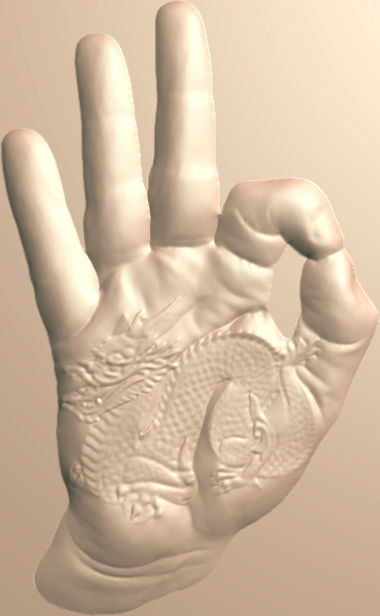}}
    \hspace{-0.12in}
  \subfigure[]{
    \label{fig:hand_dragon:d} 
    \includegraphics[angle=0,width=0.820in]{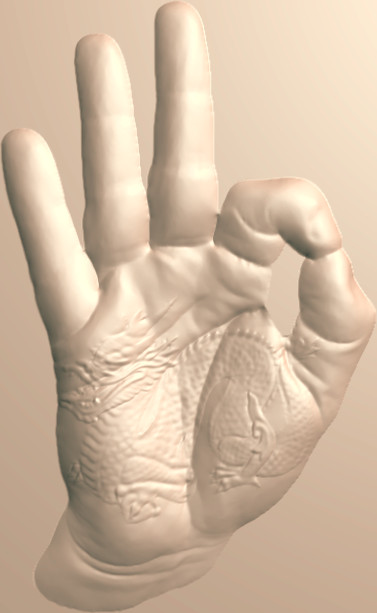}}
  \caption{\label{fig:hand_dragon}
Comparison of normal layer overlapping with normal transferring:
  (a) a dragon layer overlaps a hand layer;
  (b) the dragon layer transfers onto the hand layer;
  (c) the resulting bas-relief of (a);
  and (d) the resulting bas-relief of (b).
}
\end{figure}

In Figure \ref{fig:hand_dragon}, another example is given to demonstrate the different visual effects
between normal layer overlapping and detail transferring.
The cut-and-paste operation proposed in \cite{JMS14} induces a result with the detail layer overlapping
the hand model (see Figure \ref{fig:hand_dragon:a} and \ref{fig:hand_dragon:c}).
Our detail transferring operation generates a more reasonable blending effect, with the detail layer planted along the surface of palm.

As can be seen from figures, the previous work indicates an overlap which induces inevitable distortions between the upper and lower layers,
while our method makes the detail blend with the base surface and produces a natural effect.

%%% Figure
%%%

%%% Figure
%%%

\section{Normal from A Single Image}
In this section, we describe a two-scale approach for transforming a single image to a plausible bas-relief.
Two-scale means that a base shape and fine details are constructed separately.
\subsection{Normal Computation from Image}
Given a height field $h(x,y)$ that describes a grayscale image, a unit normal $n(x,y)$ at $(x,y)$
is given by normalizing the vector $(-h_x,-h_y,1)$.
We use a Sobel filter to approximate the derivative $(dx,dy)$ in each direction.
For pixel at $(i,j)$, and $dz = {\alpha}_1 \cdot {\alpha}_2^{|(dx,dy)|}$, thus we obtain the normal vector $v=(dx, dy, dz)$.
The resulting normal image is used to extract the low-level details only,
while the base normal image will be constructed starting from few strokes.

\subsection{Costruction of Base Normal}
\begin{figure}
  \centering
  \subfigure[]{
    \label{fig:pigbase:a} 
    \includegraphics[angle=0,width=0.820in]{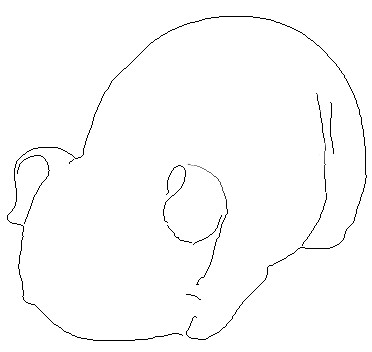}}
  \hspace{-0.12in}
  \subfigure[]{
    \label{fig:pigbase:b} 
    \includegraphics[angle=0,width=0.820in]{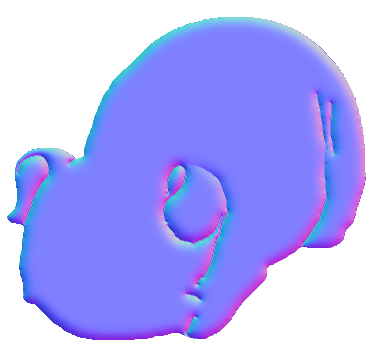}}
    \hspace{-0.12in}
  \subfigure[]{
    \label{fig:pigbase:c} 
    \includegraphics[angle=0,width=0.820in]{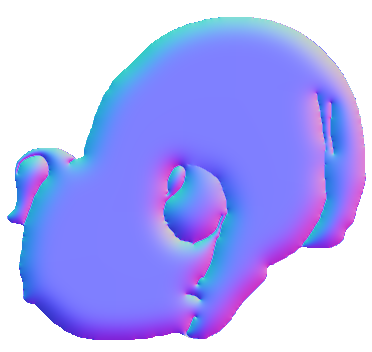}}
    \hspace{-0.12in}
  \subfigure[]{
    \label{fig:pigbase:d} 
    \includegraphics[angle=0,width=0.820in]{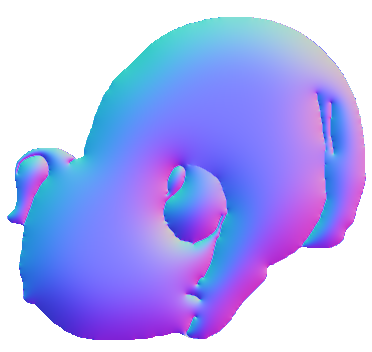}}
  \caption{\label{fig:pigbase}
An illustration of normal from strokes with different iterations:
  (a) strokes from edges computed from the image;
  (b) after 100 iterations;
  (c) after 500 iterations;
  and (d) after 1000 iterations.
}
\end{figure}

In this section, we introduce our method for the construction of a base normal image
starting from a sketch image with several strokes.
We first extract an edge map $e(u,v)$ via a canny edge detector,
then calculate its gradients $\nabla e$ which induces a gradient vector field
$\bm{g}(u,v)=(x(u,v),y(u,v))$ through minimizing the below energy functional,

\begin{equation}{\label{eqn:basenormal}}
\min_{\bm{g}(u,v)}\int_{\Omega }^{} ((\begin{Vmatrix}
\nabla x \end{Vmatrix}^2 + \begin{Vmatrix}
\nabla y \end{Vmatrix}^2) + \omega \begin{Vmatrix} \nabla e\end{Vmatrix}^2 \begin{Vmatrix}
\bm{g} - \nabla e
\end{Vmatrix}^2) dudv.
\end{equation}
The motivation behind this energy functional is that
the second energy term dominates the integrand near edges which preserves the edges,
and the energy is dominated by partial derivatives of the vector field which yields a smooth field
away from the edges.
The parameter $\omega$ is used to balance between the two terms.
This energy functional is similar to the one of the gradient vector flow \cite{XuP98}.
Once the gradient vector field $\bm{g}(u,v)$ is obtained,
a normal image is constructed by normalizing the vector at each pixel $(-x(u,v), -y(u,v), z)$,
where $z$ can be selected in a similar way described in the previous section.
In fact, we set $z$ as a constant for all $(u,v)$ to easily control the base normal image,
a small $z$ producing a lifting effect.
The solution of Equation (\ref{eqn:basenormal}) is obtained by solving two discretization linear systems of its Euler-Lagrange equations.
To be able to produce base normal images with various visual effects,
we solve them in an iterative way.
As can be seen from Figure \ref{fig:pigbase}, our method diffuses the gradient vectors of a gray-level edge image to other regions.
Relatively small number of iterations only diffuse the regions near the edges and produce a flatten result,
while large number of iterations will diffuse to the center of the figure and produce a plumpy result.
%%% Figure
%%%

\begin{figure}
  \centering
  \subfigure[]{
    \label{fig:hulu:a}
    \includegraphics[angle=0,width=0.65in]{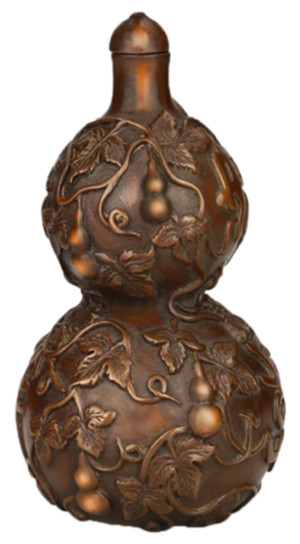}}
  \hspace{-0.1in}
  \subfigure[]{
    \label{fig:hulu:b}
    \includegraphics[angle=0,width=0.65in]{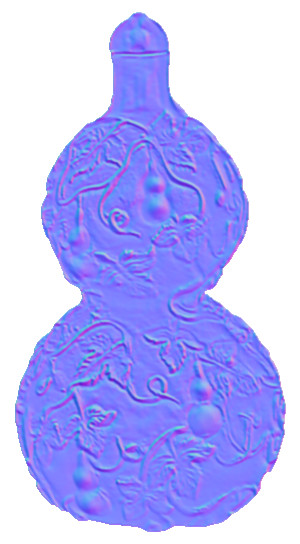}}
    \hspace{-0.1in}
  \subfigure[]{
    \label{fig:hulu:c}
    \includegraphics[angle=0,width=0.65in]{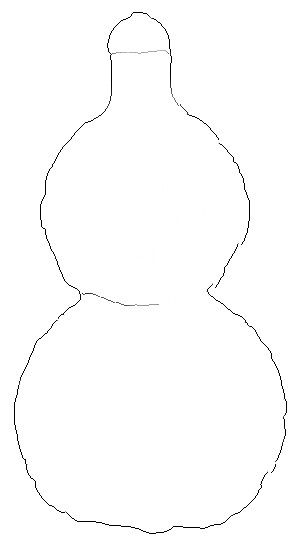}}
    \hspace{-0.1in}
  \subfigure[]{
    \label{fig:hulu:c}
    \includegraphics[angle=0,width=0.65in]{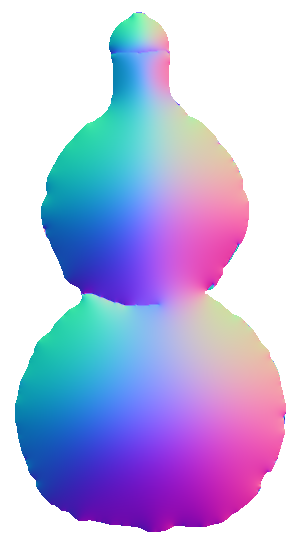}}
    \hspace{-0.1in}
  \subfigure[]{
    \label{fig:hulu:d}
    \includegraphics[angle=0,width=0.65in]{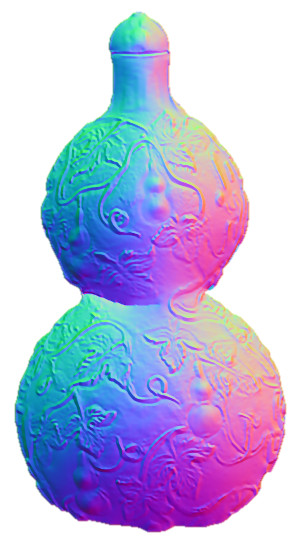}}
  \caption{\label{fig:hulu}
An example for bas-relief generation from a single RGB image:
  (a) a RGB image;
  (b) the detail normal image from (a);
  (c) a sketch image with few strokes;
  (b) the base normal image constructed from (c);
  and (d) the final composite normal image.
}
\end{figure}

Figure \ref{fig:hulu} illustrates the procedure of our method
for bas-relief generation from a general RGB image.
Our method takes a RGB image as input, converts it into a grayscale image,
calculates the normal, and then extracts a detail normal image.
We extract edges of the input image using a Canny edge detector,
then erase some edges to obtain a sketch image with a few important strokes.
A base normal image is constructed from the sketch image by minimizing equation (\ref{eqn:basenormal}).
Finally, we blend the base and detail normal images by means of the composition operation to generate a normal image of bas-relief.

%%% Figure
%%%

%%% Figure
%%%
\begin{figure}
  \centering
  \subfigure[]{
    \label{fig:pig:a} 
    \includegraphics[angle=0,width=0.820in]{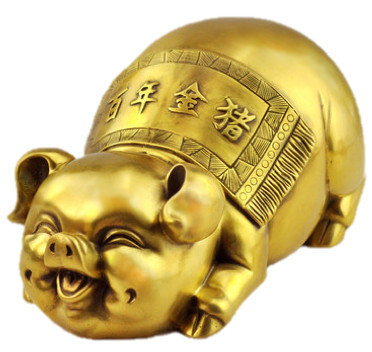}}
  \hspace{-0.12in}
  \subfigure[]{
    \label{fig:pig:b} 
    \includegraphics[angle=0,width=0.820in]{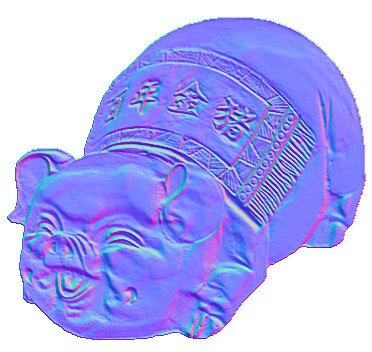}}
    \hspace{-0.12in}
  \subfigure[]{
    \label{fig:pig:c} 
    \includegraphics[angle=0,width=0.820in]{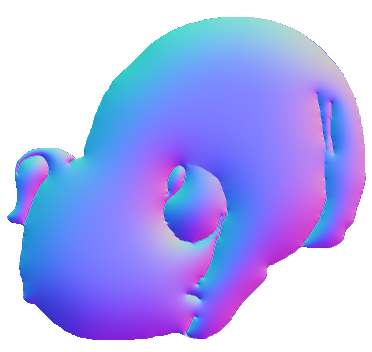}}
    \hspace{-0.12in}
  \subfigure[]{
    \label{fig:pig:d} 
    \includegraphics[angle=0,width=0.820in]{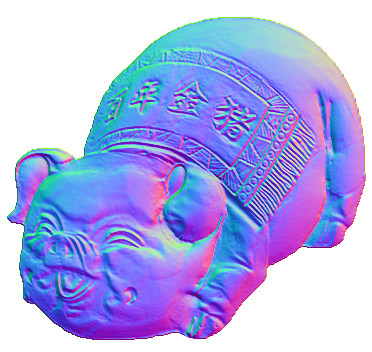}}
  \caption{\label{fig:pig}
Normal image generation from a single RGB image.
  (a) A RGB image;
  (b) the detail normal image from (a);
  (c) the base normal image;
  and (d) the composite normal image.
}
\end{figure}

Figure \ref{fig:pig} shows another example.
It is worthy to note that our construction method for base normal image is suitable with disconnected strokes.
Unlike previous related work,
our method builds the base surface from the viewpoint of gradients in a diffusion way,
so it does not rely on the boundary curves.

It is important to note that we are not proposing an algorithm to reconstruct the precise geometry with a single image. We focus on a two-scale approach to transform a single image into a plausible bas-relief with minimal efforts.

\section{Layered Stylization}

\subsection{Variational Formulation}
To expand the shape space of resulting bas-reliefs,
we attempt to extend the objective beyond the standard formulation such as in the work \cite{WDBRF07},
by introducing an additional requirement on the unknown height $z(u,v)$.
Actually, to represent a smooth base shape or a globally layered shape,
we introduce an auxiliary function into the optimization model.
Consequently, in the continuous setting, the energy minimization problem is formulated as follows,
\begin{equation}{\label{eqn:variational}}
\min_{z(u,v)}\int_{\Omega }^{}(\begin{Vmatrix}
\nabla z(u,v) - \bm{g}(u,v)
\end{Vmatrix}^2 + \lambda^2 \begin{Vmatrix}
z(u,v) - h(u,v)
\end{Vmatrix}^2) dudv,
\end{equation}
where $h(u,v)$ is a function to be as close as possible,
and $\lambda$ is used to balance these two energy terms.
The function $z(u,v)$ that minimizes this functional satisfies the Euler-Lagrange equation:
\begin{equation}{\label{eqn:screened}}
(\Delta - \lambda^2)z = \nabla \cdot \bm{g}-\lambda^2 h,
\end{equation}
which is called as screened Poisson equation.
In our framework, the second term of Equation (\ref{eqn:variational}) is treated as
a regularization term.
The $\lambda$ is a constant that controls the trade-off between the fidelity of $z(u,v)$ to the input gradient field versus the regularization term.
In our experiments, $\lambda$ is usually set to values ranging from $0$ to $1$,
higher values corresponding to a stronger bias towards the function $h(u,v)$.

\subsection{Gradient Computation}
\begin{figure}
  \centering
  \subfigure[]{
    \label{fig:torus:a} 
    \includegraphics[angle=0,width=0.820in]{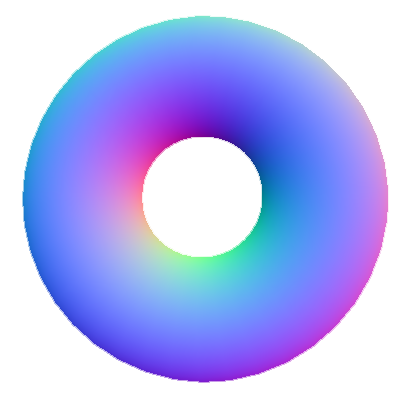}}
  \hspace{-0.12in}
  \subfigure[]{
    \label{fig:torus:b} 
    \includegraphics[angle=0,width=0.820in]{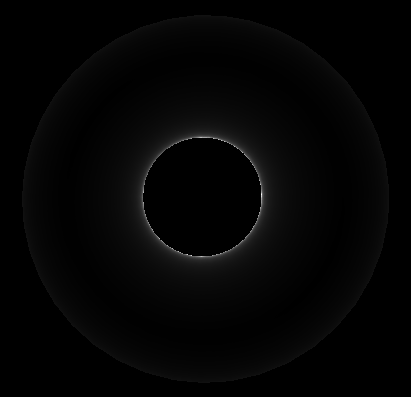}}
    \hspace{-0.12in}
  \subfigure[]{
    \label{fig:torus:c} 
    \includegraphics[angle=0,width=0.820in]{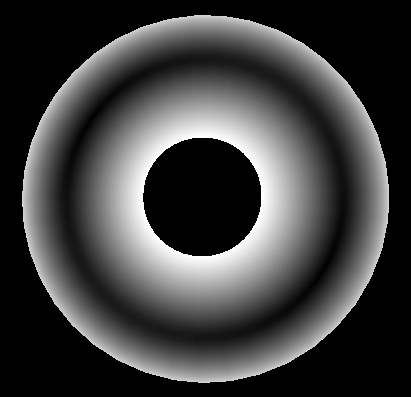}}
    \hspace{-0.12in}
  \subfigure[]{
    \label{fig:torus:d} 
    \includegraphics[angle=0,width=0.820in]{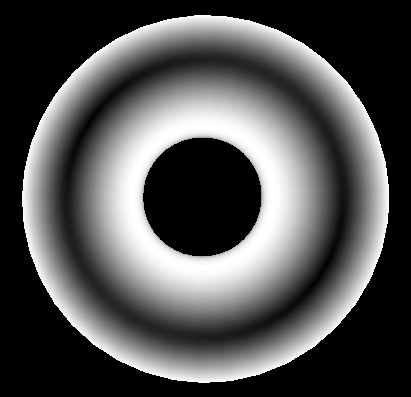}}
  \caption{\label{fig:torus}
Gradient adjustment using the $z-$component of normals $N_z$:
  (a) a given normal image,
  (b) computed gradients of (a) directly,
  (c) adjusted gradients using $\alpha=0$;
  and (d)adjusted gradients using $\alpha=1$.
}
\end{figure}

Given a normal image $N(p)=(N_x(p), N_y(p), N_z(p))$,
the gradient at a pixel $p$ is computed as
$\bm{g}(p) = ( \frac{N_x(p)}{N_z(p)}, \frac{N_y(p)}{N_z(p)} )$ in the work \cite{JMS14}.
To improve the computational stability especially when $N_z(p)$ is close to zero,
we compute the gradient $\bm{g}(p)$ at a pixel $p$ as follows,
\begin{equation}{\label{eqn:lap}}
\bm{g}(p) = ( {N_x(p)}, {N_y(p)} ) \cdot {N_z(p)}^{\alpha},
\end{equation}
where the parameter $\alpha \geq 0$ is used to reduce the gradient magnitude,
equivalently to attenuate the features along the the steep boundaries and occlusion areas where $N_z(p)$ is usually small.
Generally, a larger value of $\alpha$ generates a bas-relief with a more
flattened appearance because the corresponding magnitudes of resulting gradients fall into a smaller
range and reach uniform.
As can be seen from Figure \ref{fig:torus}, the gradients near boundaries possess relative large magnitudes (a),
and a big $\alpha$ will equalize the gradient magnitudes effectively.
In our experiment, we usually set $\alpha=1$ by default.

%%% Figure
%%%

%-------------------------------------------------------------------------
\section{Experimental Results}
In this section, we demonstrate practical applications of the proposed bas-relief modeling tool.
We start with the implementation and several experimental examples,
then compare performance to several state-of-the-art methods.

In our case, we solve the corresponding screened Poisson equation subject to Dirichlet boundary conditions.
Thus, the reconstruction of the height map from variational formulation requires solving
a linear sparse system $A \bm{X} = \bm{b}$ deduced from the discrete version of Equation (\ref{eqn:screened}).
Specifically, we use the boundary conditions wherever the relief boundary
and silhouettes between scene elements and the background correspond to zero background,
resulting in a perfectly flat background.

We now discuss the effects of various parameter settings,
especially for the key parameters $\sigma_c$ and $\lambda$.

\begin{figure}
  \centering
  \subfigure[]{
    \label{fig:arm:a} 
    \includegraphics[angle=0,width=1.0in]{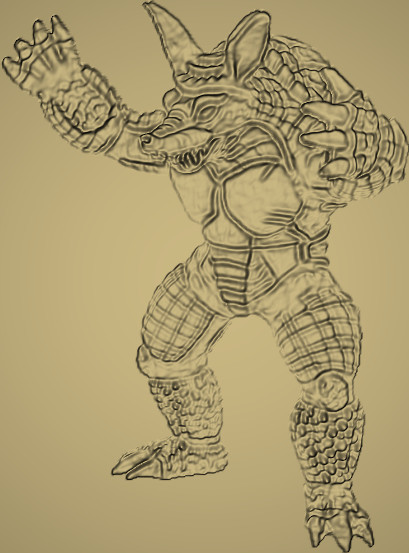}}
  \hspace{0.01in}
  \subfigure[]{
    \label{fig:arm:b} 
    \includegraphics[angle=0,width=1.0in]{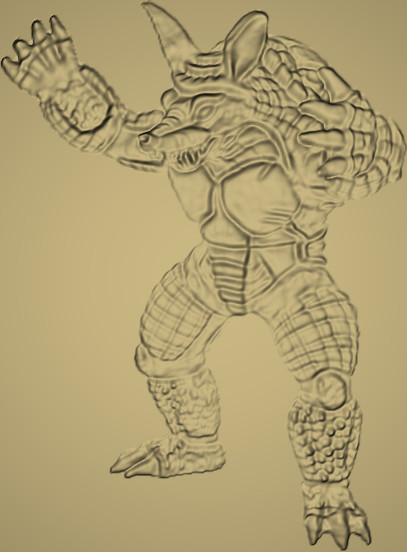}}
    \hspace{0.01in}
  \subfigure[]{
    \label{fig:arm:c} 
    \includegraphics[angle=0,width=1.0in]{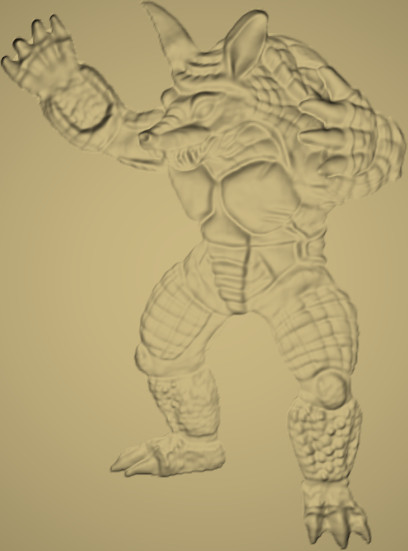}}
  \caption{\label{fig:arm}
Bas-reliefs produced using different $\sigma_c$:
  (a) $\sigma_c = 2$;
  (b) $\sigma_c = 4$;
  and (c) $\sigma_c = 8$.
}
\end{figure}

\begin{figure}
  \centering
  \subfigure[]{
    \label{fig:dragon:a} 
    \includegraphics[angle=0,width=1.0865in]{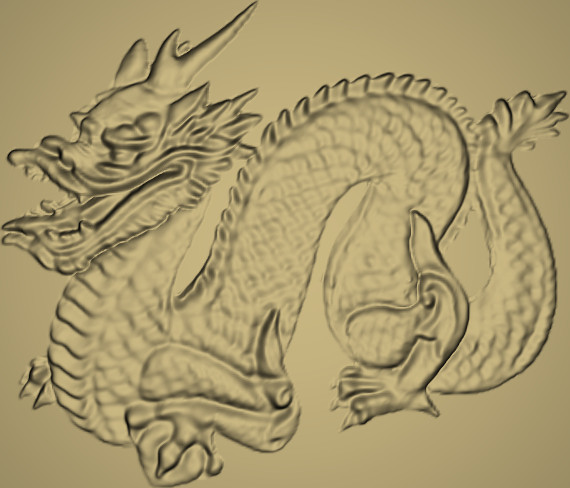}}
  \hspace{-0.08in}
  \subfigure[]{
    \label{fig:dragon:b} 
    \includegraphics[angle=0,width=1.078in]{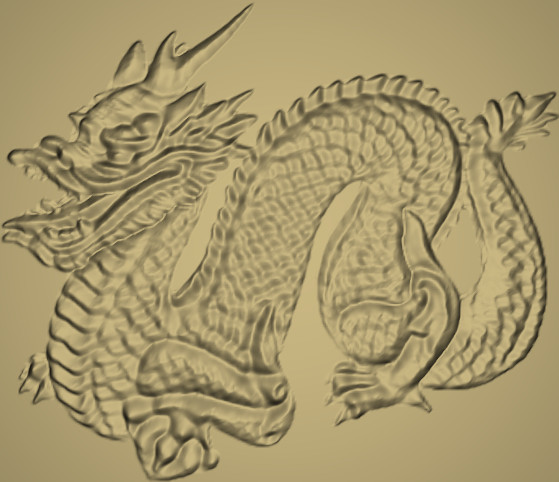}}
    \hspace{-0.08in}
  \subfigure[]{
    \label{fig:dragon:c} 
    \includegraphics[angle=0,width=1.081in]{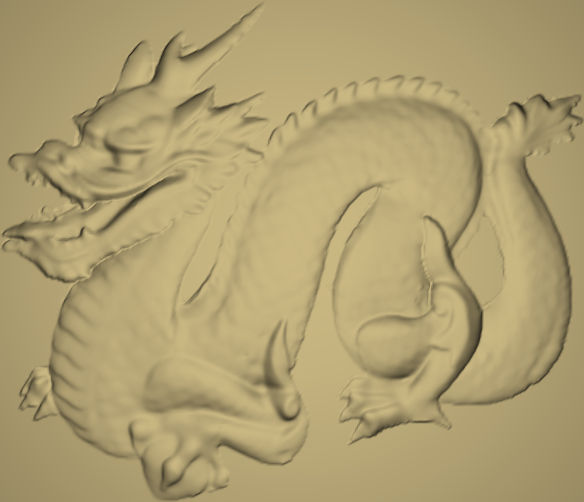}}
  \caption{\label{fig:dragon}
Detail tuning through different $\beta$ and $\gamma$:
  (a) the resulting bas-relief generated from a detail image;
  (b) the resulting bas-relief with parameters $\beta=1$ and $\gamma=1/2$;
  and (c) the resulting bas-relief with parameters $\beta=1/2$ and $\gamma=2$.
}
\end{figure}

We first emphasize the parameters which directly influence the fine details.
To analyze the parameter $\sigma_c$ in Equation (\ref{eqn:nfilter}),
$\lambda$ is set to zero in those experiments.
In Figure \ref{fig:arm}, we demonstrate the different effects of various parameters $\sigma_c$.
After integration, we linearly scale the resulting height map into the same range of depths for these three figures.
As can be seen in Figure \ref{fig:arm}, a small $\sigma_c$, corresponding to a narrow band-pass width, captures small local details well while neglects the global shape,
and a large $\sigma_c$ gathers broad bands of frequencies to capture the global shape and
produces results with similar appearance of the original model.
%%% Figure
%%%
In Figure \ref{fig:dragon}, we demonstrate the effects by rotating normal vectors to edit the extracted details. The one can enhance or attenuate the details by tuning $\beta$ and $\gamma$ easily.
It enhances the details with a big $\beta$ and a small $\gamma$ (see Figure \ref{fig:dragon:b}),
and vice versa (see Figure \ref{fig:dragon:c}).
%%% Figure
%%%

\begin{figure}
  \centering
  \subfigure[]{
    \label{fig:budda2:a}
    \includegraphics[angle=0,width=0.808in]{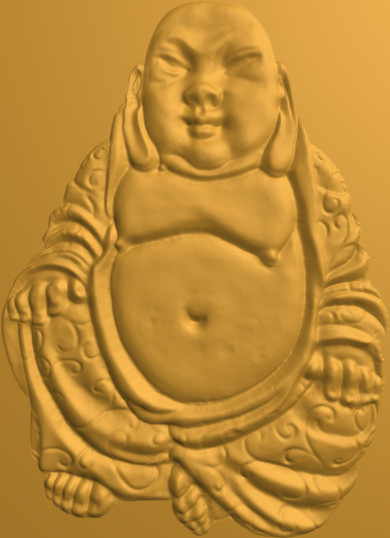}}
  \hspace{-0.09in}
  \subfigure[]{
    \label{fig:budda2:b}
    \includegraphics[angle=0,width=0.8005in]{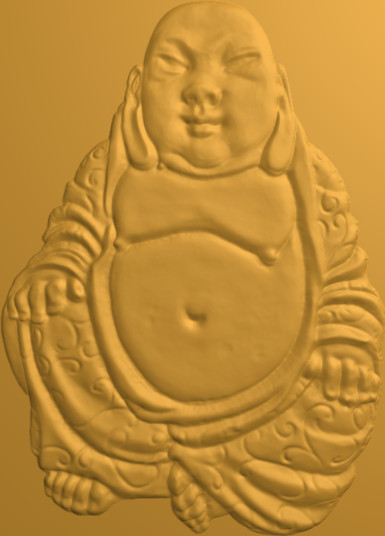}}
    \hspace{-0.09in}
  \subfigure[]{
    \label{fig:budda2:c}
    \includegraphics[angle=0,width=0.80in]{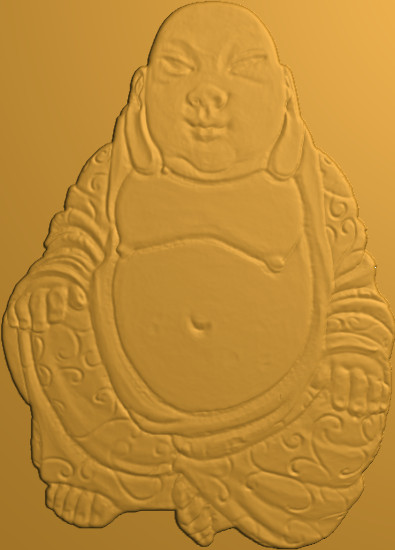}}
    \hspace{-0.09in}
  \subfigure[]{
    \label{fig:budda2:d}
    \includegraphics[angle=0,width=0.80in]{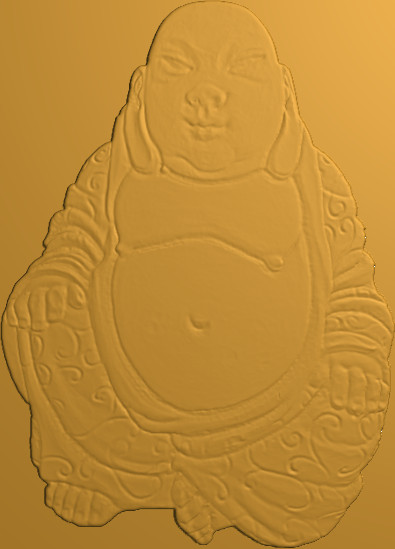}}
  \caption{\label{fig:budda2}
Illustration of effects of different $\lambda$ for the same input normal image:
  (a) $\lambda = 0$;
  (b) $\lambda = 0.1$;
  (c) $\lambda = 0.5$;
  and (d) $\lambda = 1$.
}
\end{figure}

In Figure \ref{fig:budda2}, we now demonstrate different effects by altering the parameter $\lambda$ in Equation (\ref{eqn:variational}).
For a given normal image (Figure \ref{fig:budda1:d}), our method generates a sequence of bas-reliefs ordered according to increasing $\lambda$ as shown in Figure \ref{fig:budda2}.
We set $h(u,v)$ be a constant function in this example.
As can be seen from these figures, a larger $\lambda$ exhibits more biases towards the second term
and consequently produces more flattened appearance.
%%% Figure
%%%

\begin{figure}
  \centering
  \subfigure[]{
    \label{fig:head:a} 
    \includegraphics[angle=0,width=0.80in]{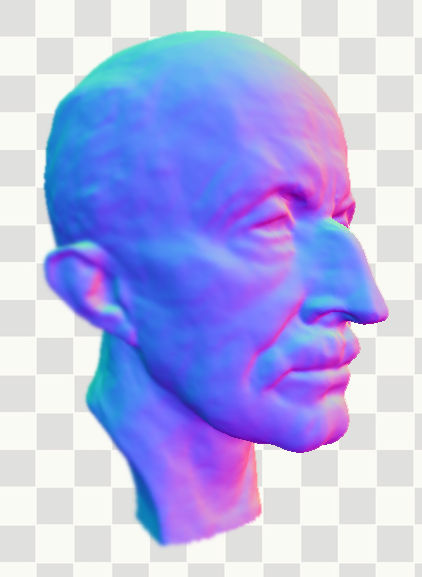}}
  \hspace{-0.092in}
  \subfigure[]{
    \label{fig:head:b} 
    \includegraphics[angle=0,width=0.82in]{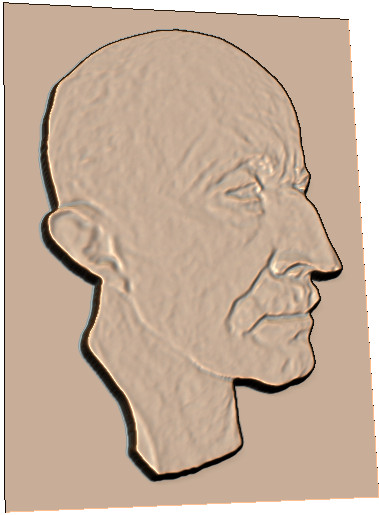}}
    \hspace{-0.092in}
  \subfigure[]{
    \label{fig:head:c} 
    \includegraphics[angle=0,width=0.80in]{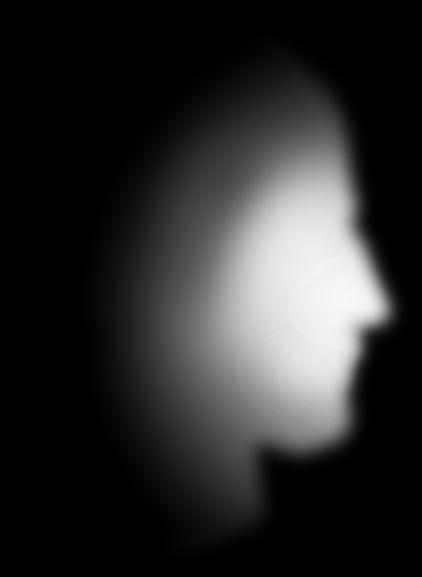}}
    \hspace{-0.092in}
  \subfigure[]{
    \label{fig:head:d} 
    \includegraphics[angle=0,width=0.82in]{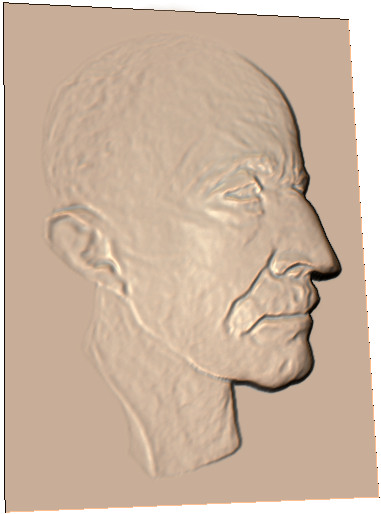}}
  \caption{\label{fig:head}
Illustration of effects of different auxiliary functions $h(u,v)$:
  (a) a normal image;
  (b) the bas-relief generated using a constant function $h(u,v)$;
  (c) a smooth $h(u,v)$;
  and (d) the bas-relief generated using (c).
}
\end{figure}

\begin{figure}
  \centering
  \subfigure[]{
    \label{fig:golf:a}
    \includegraphics[angle=0,width=0.770in]{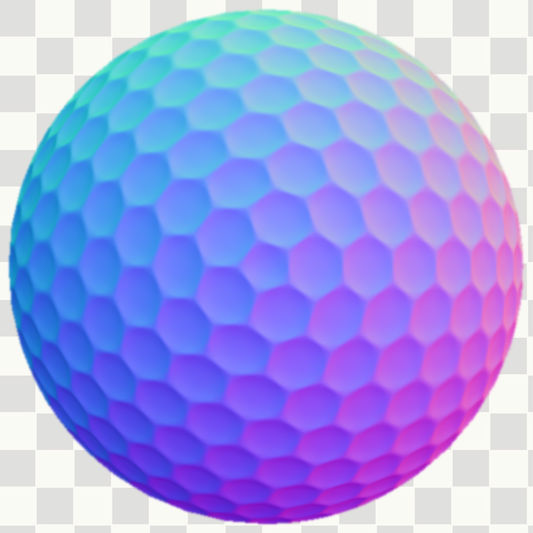}}
  \hspace{-0.085in}
  \subfigure[]{
    \label{fig:golf:b}
    \includegraphics[angle=0,width=0.838in]{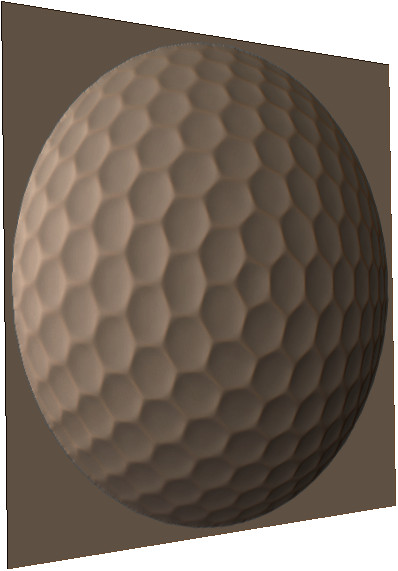}}
    \hspace{-0.085in}
  \subfigure[]{
    \label{fig:golf:c}
    \includegraphics[angle=0,width=0.77in]{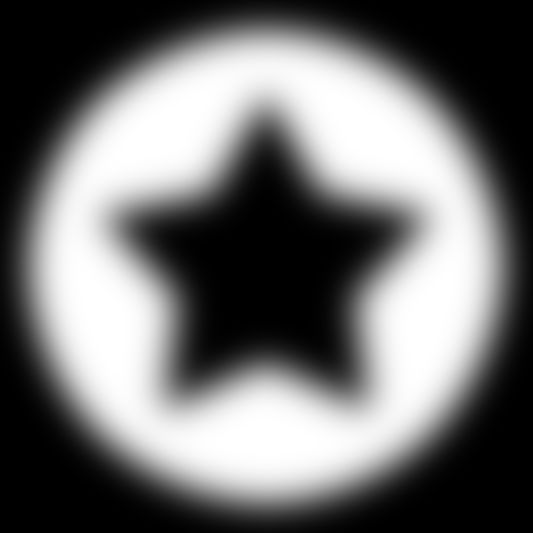}}
    \hspace{-0.085in}
  \subfigure[]{
    \label{fig:golf:d}
    \includegraphics[angle=0,width=0.838in]{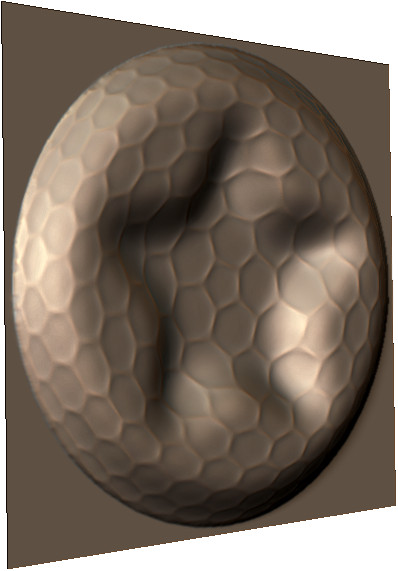}}
  \caption{\label{fig:golf}
Illustration of effects of with and without an auxiliary function $h(u,v)$:
  (a) a normal image;
  (b) the bas-relief generated without $h(u,v)$;
  (c) a smooth function $h(u,v)$;
  and (d) the bas-relief generated using (c).
}
\end{figure}

We now discuss the application of the auxiliary function $h(u,v)$ in Equation (\ref{eqn:variational}).
As a matter of fact, $h(u,v)$ in our context can be treated as a base surface which can be a smooth or step-shaped function.
Figure \ref{fig:head} shows an example where we consider the effect of a base surface with a gradual change from left to right (Figure \ref{fig:head:c}).
Compared with Figure \ref{fig:head:b}, Figure \ref{fig:head:d} exhibits different feelings of stereo perception. Another example of smooth function $h(u,v)$ is shown in Figure \ref{fig:golf}.
A pentagram is imposed on a semisphere to generate an interesting bas-relief.
%%% Figure
%%%
%%% Figure
%%%

\begin{figure}
  \centering
  \subfigure[]{
    \label{fig:grape:a} 
    \includegraphics[angle=0,width=0.85in]{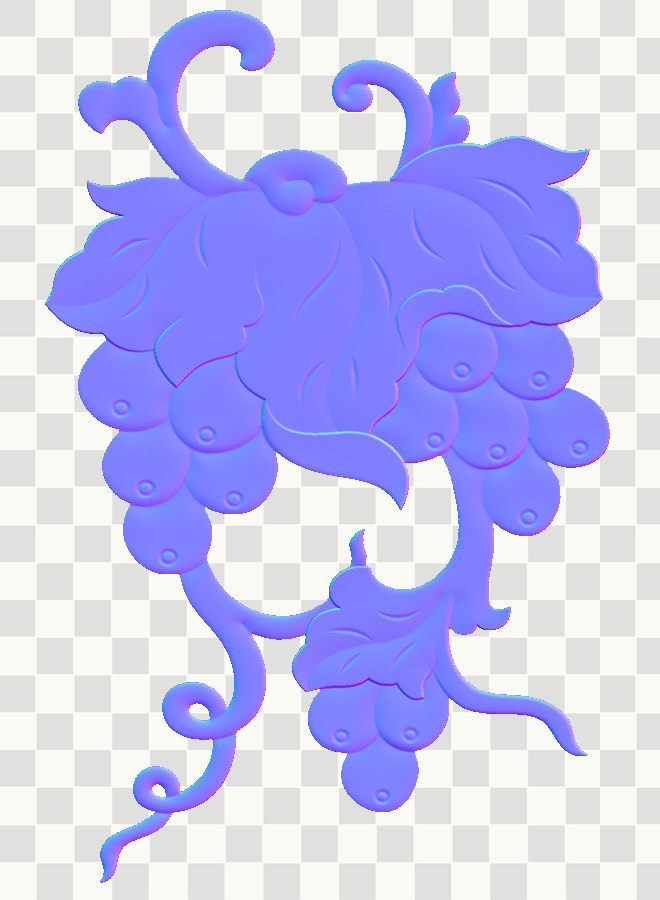}}
  \hspace{-0.0782in}
  \subfigure[]{
    \label{fig:grape:b} 
    \includegraphics[angle=0,width=0.75in]{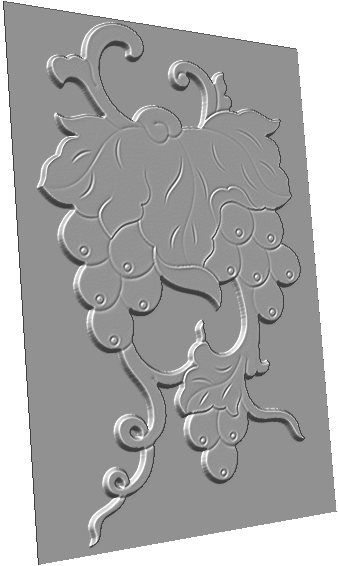}}
    \hspace{-0.0782in}
  \subfigure[]{
    \label{fig:grape:c} 
    \includegraphics[angle=0,width=0.85in]{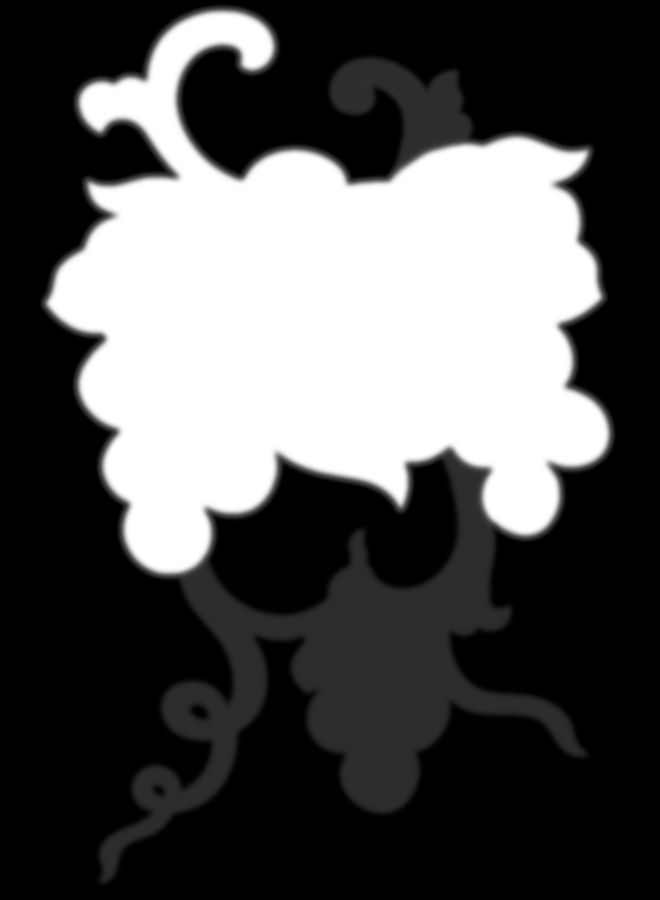}}
    \hspace{-0.0782in}
  \subfigure[]{
    \label{fig:grape:d} 
    \includegraphics[angle=0,width=0.75in]{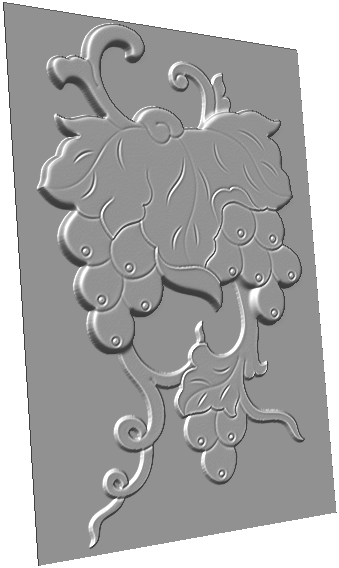}}
  \caption{\label{fig:grape}
Illustration of effects of different auxiliary function $h(u,v)$:
  (a) a normal image;
  (b) the bas-relief generated using a constant function $h(u,v)$;
  (c) a hierarchic function $h(u,v)$ with two layers;
  and (d) the bas-relief generated using (c).
}
\end{figure}

\begin{figure}
  \centering
  \subfigure[]{
\hspace{-0.1in}
    \label{fig:flower:a}
    \includegraphics[angle=0,width=1.6in]{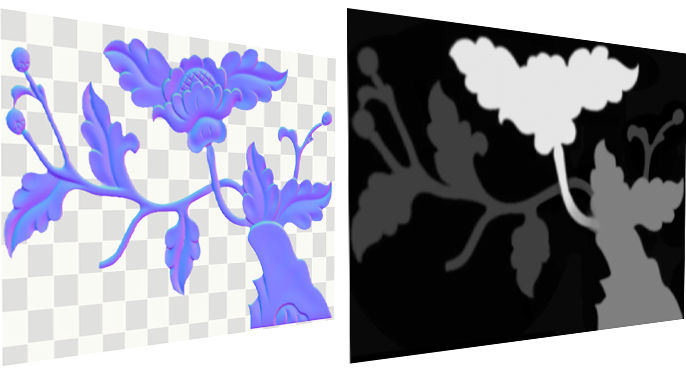}}
    \hspace{-0.02in}
  \subfigure[]{
    \label{fig:flower:b}
    \includegraphics[angle=0,width=1.66in]{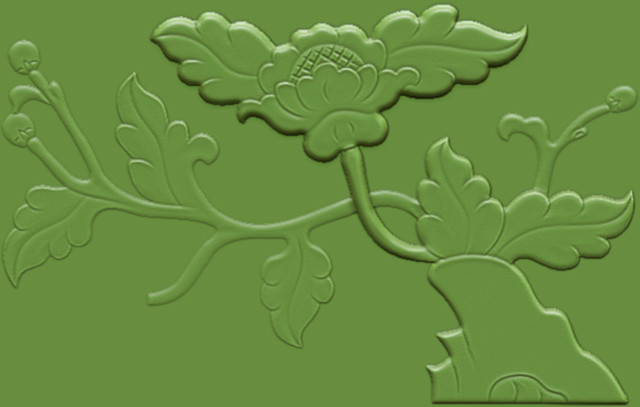}}
  \caption{\label{fig:flower}
An example with three layers:
  (a) a normal image and a hierarchic function $h(u,v)$ with three layers;
  and (c) the resulting bas-relief.
}
\end{figure}

\textbf{Layered Stylization.} Besides smoothed base surfaces, step functions can be integrated into our system to generate more levels of depth.
In this paper, we call this effect as layered stylization.
Figure \ref{fig:grape} shows an example integrated with a two-step function.
Compared with Figure \ref{fig:grape:b}, Figure \ref{fig:grape:d} offers a layered appearance.
In our implementation, one is allowed to produce more layers if necessary.
Figure \ref{fig:flower} shows an example with three layers which makes the flower stand out from the stem and leaves.
%%% Figure
%%%

%%% Figure
%%%

%%% Figure
%%%
\begin{figure}
  \centering
  \subfigure[]{
    \label{fig:sunflower:a} 
    \includegraphics[angle=0,width=0.66in]{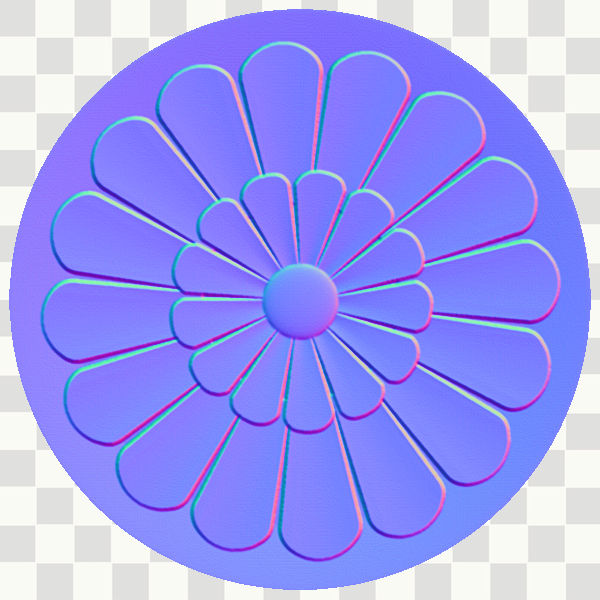}}
  \hspace{-0.11in}
  \subfigure[]{
    \label{fig:sunflower:b} 
    \includegraphics[angle=0,width=1.32in]{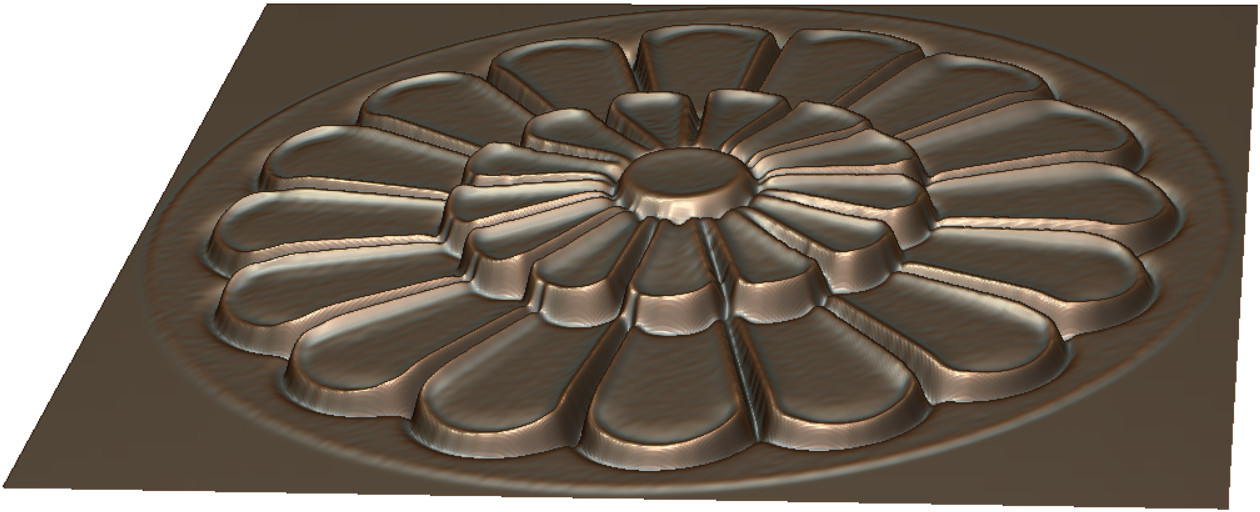}}
    \hspace{-0.12in}
  \subfigure[]{
    \label{fig:sunflower:c} 
    \includegraphics[angle=0,width=1.32in]{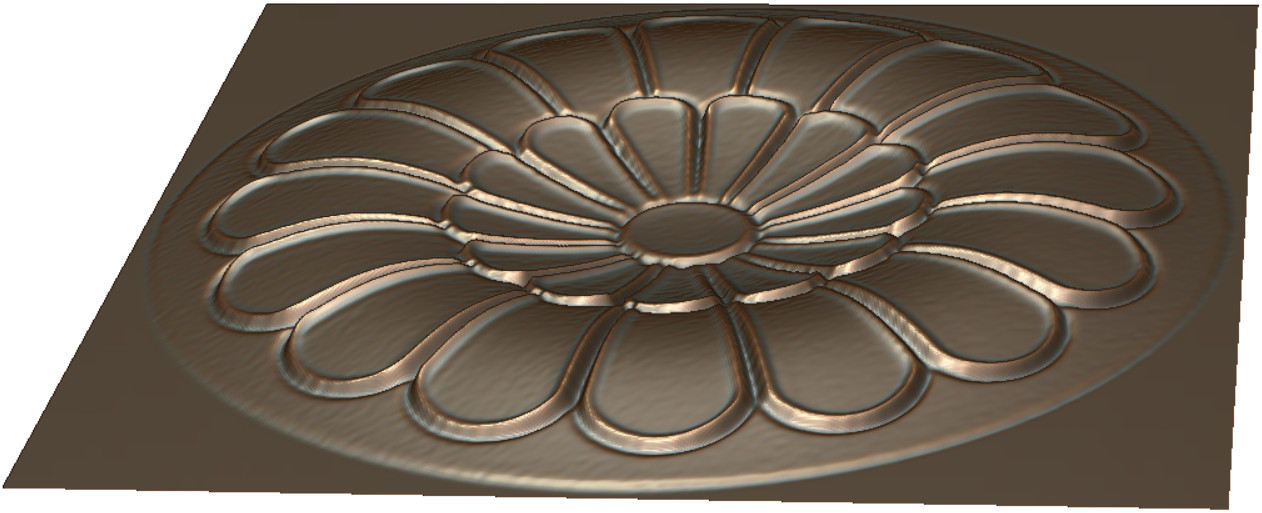}}
  \caption{\label{fig:sunflower}
Illustration of effects of different auxiliary function $h(u,v)$:
  (a) a normal image;
  (b) the bas-relief generated using a constant function $h(u,v)$;
  (c) a hierarchic function $h(u,v)$ with two layers;
  and (d) the bas-relief generated using (c).
}
\end{figure}

\begin{figure}
  \centering
  \subfigure[]{
    \label{fig:comparison:a} %
    \includegraphics[angle=0,width=0.99in]{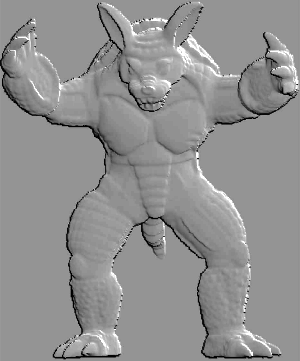}}
  \hspace{0.0in}
  \subfigure[]{
    \label{fig:comparison:b} %
    \includegraphics[angle=0,width=1.0in]{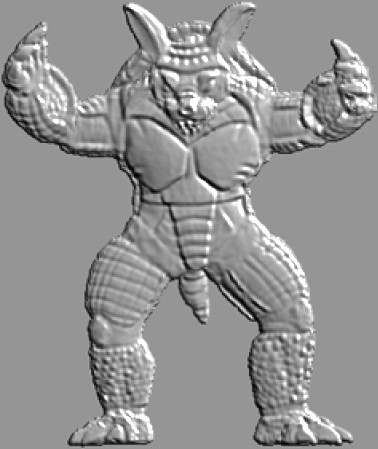}}
    \hspace{0.0in}
  \subfigure[]{
    \label{fig:comparison:c} %
    \includegraphics[angle=0,width=1.0in]{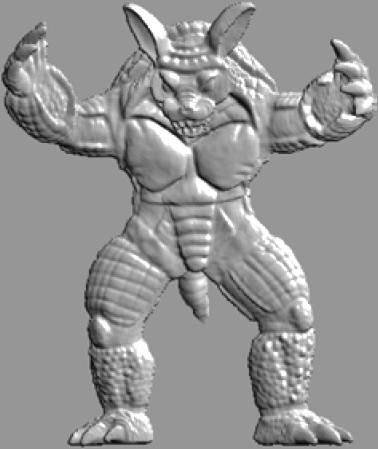}}
    \hspace{0.0in}
  \subfigure[]{
    \label{fig:comparison:d} %
    \includegraphics[angle=0,width=1.0in]{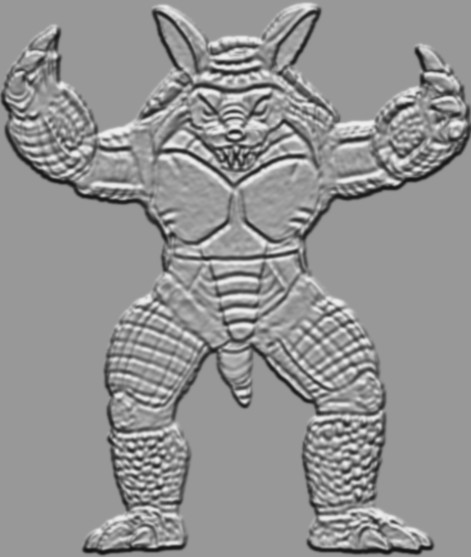}}
    \hspace{0.0in}
  \subfigure[]{
    \label{fig:comparison:e} %
    \includegraphics[angle=0,width=0.99in]{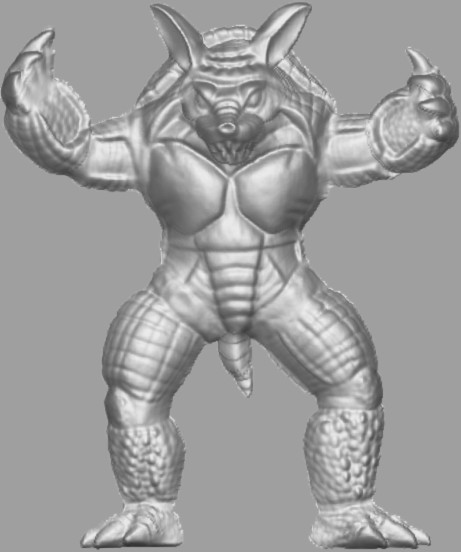}}
    \hspace{0.0in}
  \subfigure[]{
    \label{fig:comparison:f} %
    \includegraphics[angle=0,width=1.006in]{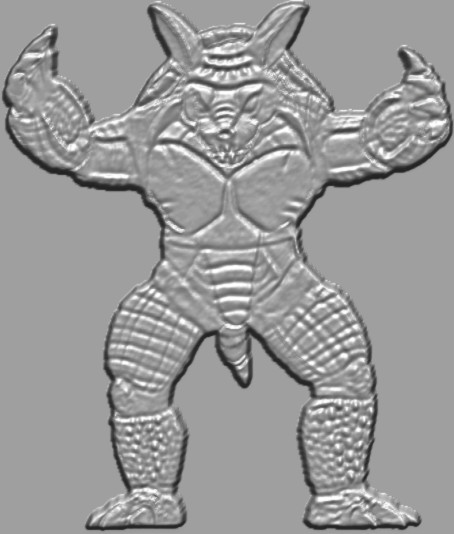}}
    \hspace{0.0in}
  \subfigure[]{
    \label{fig:comparison:g} %
    \includegraphics[angle=0,width=1.005in]{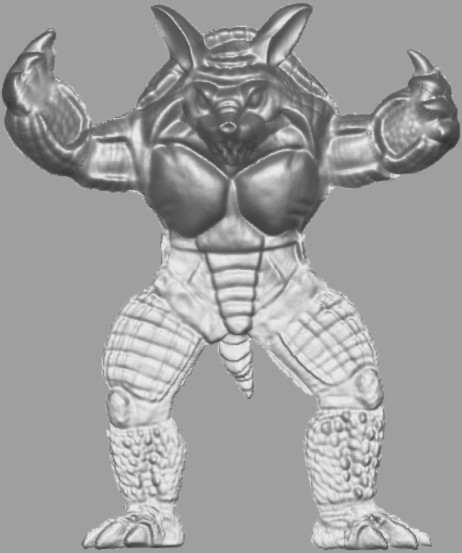}}
    \hspace{0.0in}
  \subfigure[]{
    \label{fig:comparison:h} %
    \includegraphics[angle=0,width=1.0in]{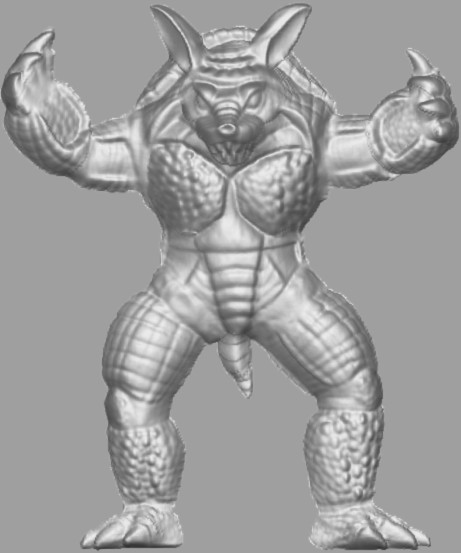}}
    \hspace{0.0in}
  \subfigure[]{
    \label{fig:comparison:i} %
    \includegraphics[angle=0,width=1.02in]{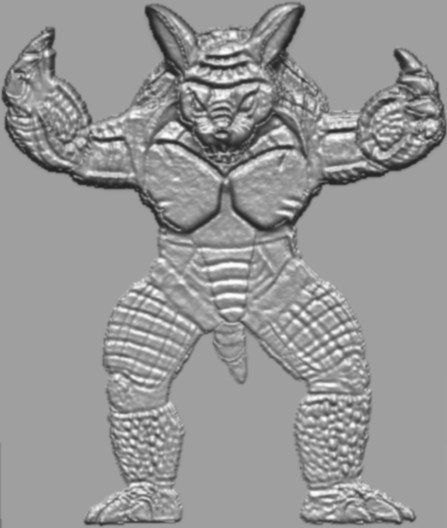}}
  \caption{\label{fig:comparison} Bas-reliefs produced by
  (a) the method of Cignoni et al. \cite{CMS97};
  (b) the method of Kerber \cite{Ker07};
  (c) the method of Sun et al. \cite{SPRF09};
  (d) the method of Ji et al. \cite{JSLW14};
  (e) our method with $\lambda = 0$;
  (f) our method with $\lambda = 1$ and a constant function $h(u,v)$;
  (g) our method with $\lambda = 0$;
  (h) our method with $\lambda = 0$;
  and (i) our method with $\lambda = 1$ and a step function $h(u,v)$.
}
\end{figure}

Now we turn to compare our results with those of some previous methods.
We begin by comparing our results with those of \cite{CMS97},
 \cite{Ker07}, \cite{SPRF09}, and \cite{JSLW14}.
Result for Kerber's approach was obtained using default parameter settings.
Result for Sun's approach was obtained using the following parameters:
$B = 10000$, $m_0 = 32$, $n = 4$, $l = 16$, and $K = 1$.
Result for Ji's approach was obtained using Scheme II and $\lambda = 1$.
Figure \ref{fig:comparison} shows results using the above mentioned approaches.
Most of these approaches produced natural or feature-enhanced results.
Our method can be seen as an extension of the method proposed by Ji et al. \cite{JMS14}.
On one hand, both methods take normal images as inputs.
As can be seen from Figure \ref{fig:comparison:e}-\ref{fig:comparison:f},
our method can produce similar styles of bas-reliefs as their method.
However, our method has an ability to decompose a given normal image into
a detail layer and a base layer via the proposed DoG-like filter.
Consequently, our method is apt to create the hybrid style of bas-reliefs by combining different types of normal images,
and transfer details from one normal image to other normal images.
Figure \ref{fig:comparison:g} demonstrates a hybrid style of bas-relief which is composed of
a original normal image (the upper part of the body) and a detail normal image (the lower part of the body).
Figure \ref{fig:comparison:h} shows an example with details transferred.
Note that the bumps on legs are replicated and transferred to the chest through the normal decomposition-and-composition operations.
For the result shown in Figure \ref{fig:comparison:g} and Figure \ref{fig:comparison:h},
we set $\lambda = 0$ to preserve the input normals as much as possible.
On the other hand, our method integrates an auxiliary function $h(u,v)$ into a
variational formulation (in Equation (\ref{eqn:variational})) to further extend their method.
As shown in Figure \ref{fig:comparison:i}, we use a step function to generate a layered result.
In general, our method is capable of generating diversified results,
such as round, flattened, hybrid or even layered bas-reliefs.

%%% Figure
%%%

\begin{figure}
  \centering
  \vspace{-0.11in}
  \subfigure[]{
    \label{fig:examples:a}
    \includegraphics[angle=0,width=0.738in]{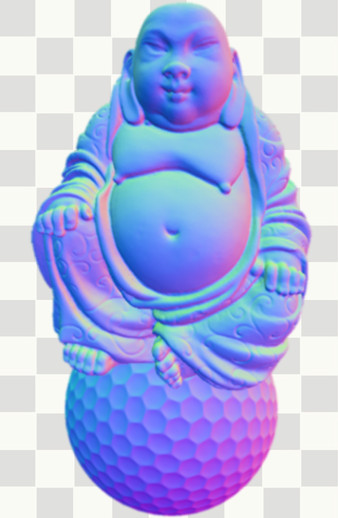}}
  \hspace{-0.023in}
 \vspace{-0.11in}
  \subfigure[]{
    \label{fig:examples:b}
    \includegraphics[angle=0,width=0.684in]{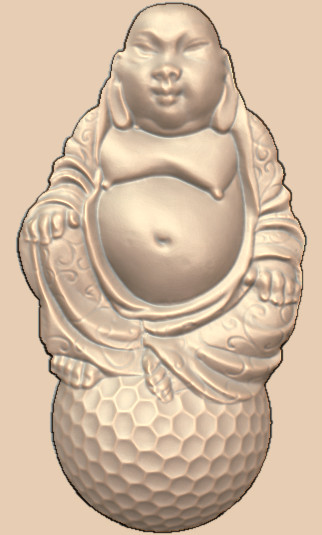}}
  \hspace{-0.023in}
  \subfigure[]{
    \label{fig:examples:c}
    \includegraphics[angle=0,width=0.738in]{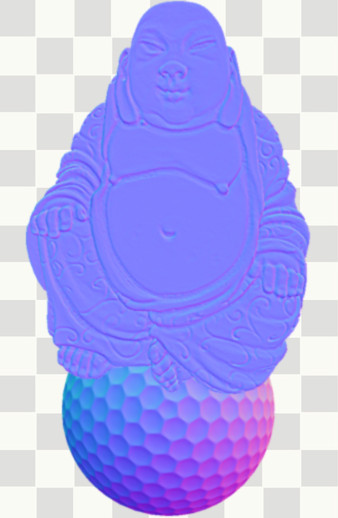}}
  \hspace{-0.023in}
  \subfigure[]{
    \label{fig:examples:d}
    \includegraphics[angle=0,width=0.6858in]{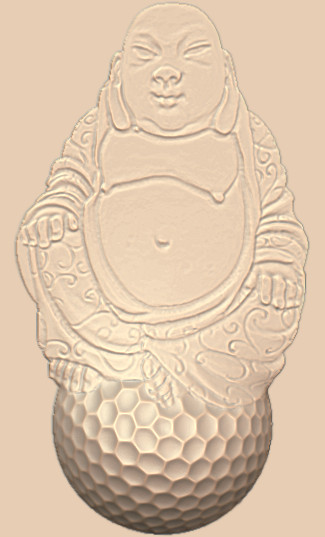}}
  \vspace{-0.11in}
  \subfigure[]{
    \label{fig:examples:e}
    \includegraphics[angle=0,width=1.0314in]{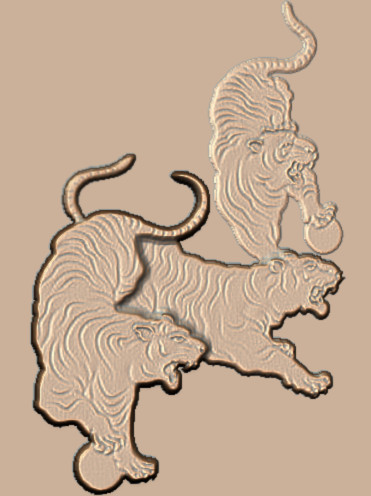}}
  \hspace{0.09in}
  \subfigure[]{
    \label{fig:examples:f}
    \includegraphics[angle=0,width=1.899in]{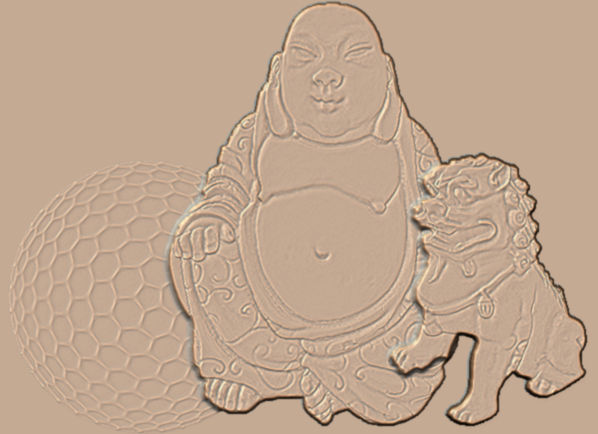}}
  \vspace{-0.11in}
  \subfigure[]{
    \label{fig:examples:g}
    \includegraphics[angle=0,width=1.107in]{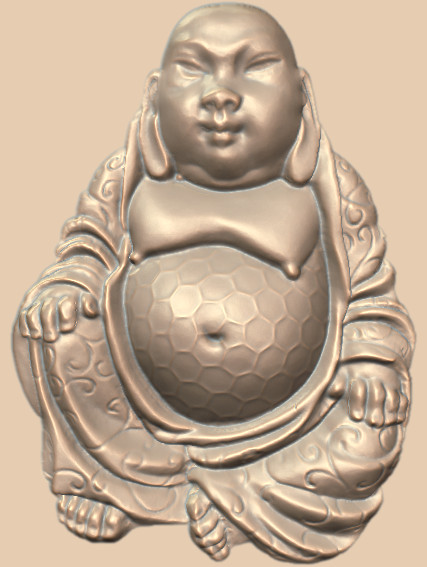}}
  \hspace{0.01in}
  \subfigure[]{
    \label{fig:examples:h}
    \includegraphics[angle=0,width=1.1043in]{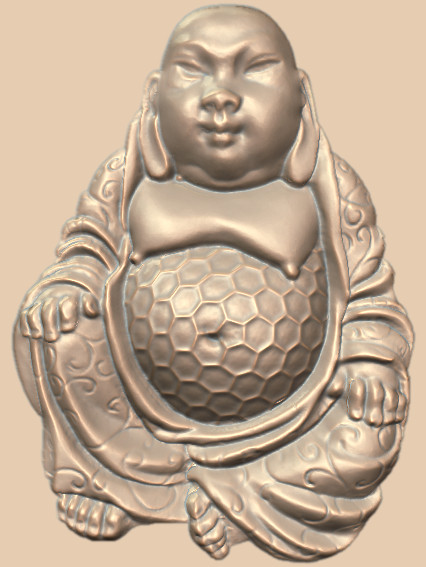}}
  \hspace{0.01in}
  \subfigure[]{
    \label{fig:examples:i}
    \includegraphics[angle=0,width=0.675in]{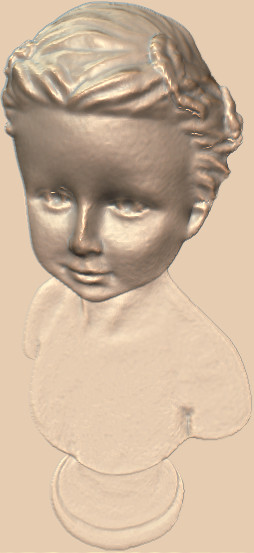}}
  \vspace{-0.11in}
  \subfigure[]{
    \label{fig:examples:j}
    \includegraphics[angle=0,width=1.692in]{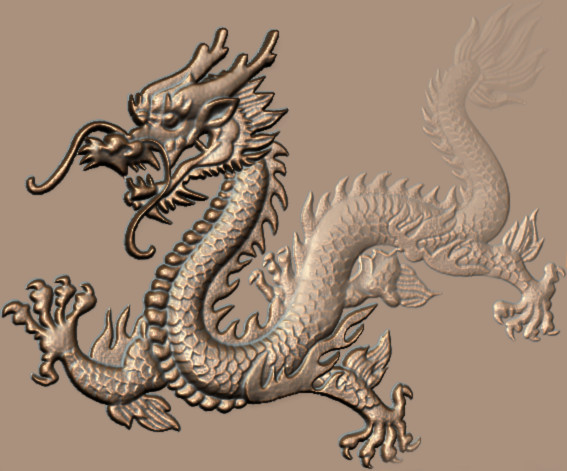}}
  \hspace{0.1in}
  \subfigure[]{
    \label{fig:examples:k}
    \includegraphics[angle=0,width=1.2348in]{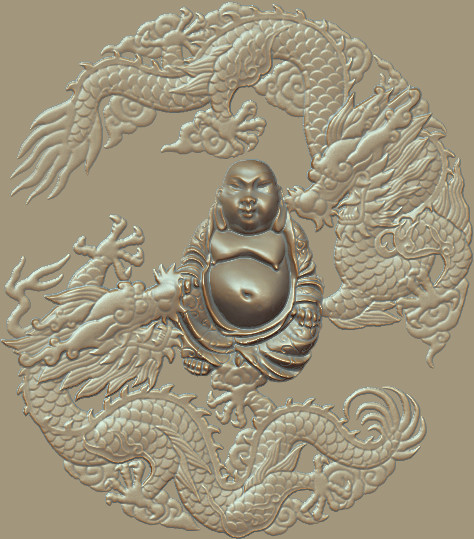}}
  \hspace{-0.04in}
  \subfigure[]{
    \label{fig:examples:l}
    \includegraphics[angle=0,width=3.17in]{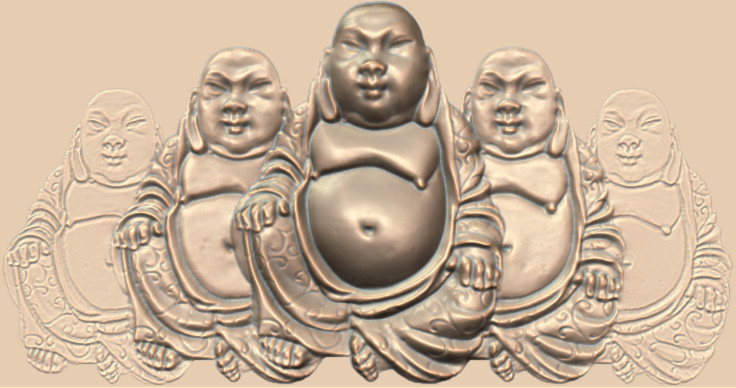}}
  \caption{\label{fig:examples} More bas-reliefs with hybrid and layered styles created using our method.
}
\end{figure}

To demonstrate the ability of our method for different styles,
we produce more bas-reliefs using our method as shown in Figure \ref{fig:examples}.
In Figure \ref{fig:examples:a}-\ref{fig:examples:d},
we demonstrate that our method is capable of creating different styles of bas-reliefs
by combining different types of normal images which are extracted using our decomposition filter.
In Figure \ref{fig:examples:e}-\ref{fig:examples:f}, we use step functions to create layered bas-reliefs.
In Figure \ref{fig:examples:g}-\ref{fig:examples:h}, we add a detail patch sampled from a golf model into the Buddha normal images by the normal composition operation.
In Figure \ref{fig:examples:h}, we use the tuning function of rotation angle (Equation (\ref{eqn:rescale})) to boost the details of golf, and then combine the patch with the Buddha by the composition operation.
In Figure \ref{fig:examples:i} and Figure \ref{fig:examples:j}, we edit base normal images partially, and then combine them with the accompanying detail normal images to yield bas-reliefs with impressive effects.
Figure \ref{fig:examples:k} and Figure \ref{fig:examples:l} demonstrate two examples created by combining various normal images which are extracted through different bands using our decomposition filter.
%%% Figure
%%%

\section{Conclusions and Future Work}
This paper focuses on extending the normal based bas-relief modeling method.
To edit a normal image more reasonably, we develop decomposition-and-composition operations on normal images.
Our approach allows for changing the visually different appearances of the resulting bas-reliefs
by editing the normal image in a layer-based way and by integrating an auxiliary function into the variational framework.
The proposed method improves the bas-relief stylization and expands the bas-relief shape space which includes hybrid and layered styles.
Numerous examples show that our method can generate reasonable and pleasant bas-reliefs.

Our work can further enrich the research topic of bas-reliefs.
One of our future work is to decompose the normal image using more sophisticated techniques such as wavelet theory, which can decompose and reconstruct the image in a multi-scale fashion.
Another future work focuses on methodologies for constructing sharp features from the perspective of normals.
A straightforward method is to automatically detect feature curves or to manually indicate some strokes near the candidates.
We can then reassign normals along feature lines to construct sharp features explicitly.
In addition, how to generate the auxiliary function automatically from a normal image or a depth image of a 3D scene is an intriguing problem deserving further investigations in the future work.

\section*{Acknowledgements}

This work was partially supported by the National Natural Science
Foundation of China (61572161,61202278), Key Laboratory of Complex Systems Modeling and Simulation (Ministry of Education, China), EPSRC (EP/J02211X/1), Research Grants Council of Hong Kong SAR (CityU 118512), and City University of Hong Kong (SRG 7004072).

\bibliographystyle{ACM-Reference-Format}
\bibliography{basrelief}


\begin{thebibliography}{00}

%%% the Web address.

\ifx \showCODEN    \undefined \def \showCODEN     #1{\unskip}     \fi
\ifx \showDOI      \undefined \def \showDOI       #1{#1}\fi
\ifx \showISBNx    \undefined \def \showISBNx     #1{\unskip}     \fi
\ifx \showISBNxiii \undefined \def \showISBNxiii  #1{\unskip}     \fi
\ifx \showISSN     \undefined \def \showISSN      #1{\unskip}     \fi
\ifx \showLCCN     \undefined \def \showLCCN      #1{\unskip}     \fi
\ifx \shownote     \undefined \def \shownote      #1{#1}          \fi
\ifx \showarticletitle \undefined \def \showarticletitle #1{#1}   \fi
\ifx \showURL      \undefined \def \showURL       {\relax}        \fi
% invisible to TeX
\providecommand\bibfield[2]{#2}
\providecommand\bibinfo[2]{#2}
\providecommand\natexlab[1]{#1}
\providecommand\showeprint[2][]{arXiv:#2}

\bibitem[\protect\citeauthoryear{Bian and Hu}{Bian and Hu}{2011}]%
        {BianH11}
\bibfield{author}{\bibinfo{person}{Zhe Bian} {and} \bibinfo{person}{Shi-Min
  Hu}.} \bibinfo{year}{2011}\natexlab{}.
\newblock \showarticletitle{Preserving detailed features in digital bas-relief
  making}.
\newblock \bibinfo{journal}{{\em Computer Aided Geometric Design\/}}
  \bibinfo{volume}{28}, \bibinfo{number}{4} (\bibinfo{year}{2011}),
  \bibinfo{pages}{245--256}.
\newblock


\bibitem[\protect\citeauthoryear{Cignoni, Montani, and Scopigno}{Cignoni
  et~al\mbox{.}}{1997}]%
        {CMS97}
\bibfield{author}{\bibinfo{person}{P. Cignoni}, \bibinfo{person}{C. Montani},
  {and} \bibinfo{person}{R. Scopigno}.} \bibinfo{year}{1997}\natexlab{}.
\newblock \showarticletitle{Computer-assisted generation of bas- and
  high-reliefs}.
\newblock \bibinfo{journal}{{\em J. Graph. Tools\/}} \bibinfo{volume}{2},
  \bibinfo{number}{3} (\bibinfo{year}{1997}), \bibinfo{pages}{15--28}.
\newblock
\showISSN{1086-7651}


\bibitem[\protect\citeauthoryear{Fattal, Lischinski, and Werman}{Fattal
  et~al\mbox{.}}{2002}]%
        {FLW02}
\bibfield{author}{\bibinfo{person}{Raanan Fattal}, \bibinfo{person}{Dani
  Lischinski}, {and} \bibinfo{person}{Michael Werman}.}
  \bibinfo{year}{2002}\natexlab{}.
\newblock \showarticletitle{Gradient domain high dynamic range compression}. In
  \bibinfo{booktitle}{{\em SIGGRAPH '02: Proceedings of the 29th annual
  conference on Computer graphics and interactive techniques}}.
  \bibinfo{pages}{249--256}.
\newblock
\showISBNx{1-58113-521-1}


\bibitem[\protect\citeauthoryear{Ji, Ma, and Sun}{Ji et~al\mbox{.}}{2014a}]%
        {JMS14}
\bibfield{author}{\bibinfo{person}{Zhongping Ji}, \bibinfo{person}{Weiyin Ma},
  {and} \bibinfo{person}{Xianfang Sun}.} \bibinfo{year}{2014}\natexlab{a}.
\newblock \showarticletitle{Bas-Relief Modeling from Normal Images with
  Intuitive Styles}.
\newblock \bibinfo{journal}{{\em IEEE Transactions on Visualization and
  Computer Graphics\/}} \bibinfo{volume}{20}, \bibinfo{number}{5}
  (\bibinfo{year}{2014}), \bibinfo{pages}{675--685}.
\newblock


\bibitem[\protect\citeauthoryear{Ji, Sun, Li, and Wang}{Ji
  et~al\mbox{.}}{2014b}]%
        {JSLW14}
\bibfield{author}{\bibinfo{person}{Zhongping Ji}, \bibinfo{person}{Xianfang
  Sun}, \bibinfo{person}{Shi Li}, {and} \bibinfo{person}{Yigang Wang}.}
  \bibinfo{year}{2014}\natexlab{b}.
\newblock \showarticletitle{Real-time Bas-Relief Generation from
  Depth-and-Normal Maps on {GPU}}.
\newblock \bibinfo{journal}{{\em Computer Graphics Forum\/}}
  \bibinfo{volume}{33}, \bibinfo{number}{5} (\bibinfo{year}{2014}),
  \bibinfo{pages}{75--83}.
\newblock


\bibitem[\protect\citeauthoryear{Kerber}{Kerber}{2007}]%
        {Ker07}
\bibfield{author}{\bibinfo{person}{Jens Kerber}.}
  \bibinfo{year}{2007}\natexlab{}.
\newblock {\em \bibinfo{title}{Digital Art of Bas-Relief Sculpting}}.
\newblock Masters thesis. \bibinfo{school}{Universit{\"a}t des Saarlandes}.
\newblock


\bibitem[\protect\citeauthoryear{Kerber, Belyaev, and Seidel}{Kerber
  et~al\mbox{.}}{2007}]%
        {kbs07}
\bibfield{author}{\bibinfo{person}{Jens Kerber}, \bibinfo{person}{Alexander
  Belyaev}, {and} \bibinfo{person}{Hans-Peter Seidel}.}
  \bibinfo{year}{2007}\natexlab{}.
\newblock \showarticletitle{Feature preserving depth compression of range
  images}. In \bibinfo{booktitle}{{\em Proceedings of the 23rd spring
  conference on computer graphics}}. Budmerice, Slovakia,
  \bibinfo{pages}{110--114}.
\newblock


\bibitem[\protect\citeauthoryear{Kerber, Tevs, Belyaev, Zayer, and
  Seidel}{Kerber et~al\mbox{.}}{2010}]%
        {KTA2010}
\bibfield{author}{\bibinfo{person}{Jens Kerber}, \bibinfo{person}{Art Tevs},
  \bibinfo{person}{Alexander Belyaev}, \bibinfo{person}{Rhaleb Zayer}, {and}
  \bibinfo{person}{Hans-Peter Seidel}.} \bibinfo{year}{2010}\natexlab{}.
\newblock \showarticletitle{Real-time Generation of Digital Bas-Reliefs}.
\newblock \bibinfo{journal}{{\em Journal of Computer-Aided Design and
  Applications\/}} \bibinfo{volume}{7}, \bibinfo{number}{4}
  (\bibinfo{year}{2010}), \bibinfo{pages}{465--478}.
\newblock


\bibitem[\protect\citeauthoryear{Kerber, Tevs, Zayer, Belyaev, and
  Seidel}{Kerber et~al\mbox{.}}{2009}]%
        {KTRAH09}
\bibfield{author}{\bibinfo{person}{Jens Kerber}, \bibinfo{person}{Art Tevs},
  \bibinfo{person}{Rhaleb Zayer}, \bibinfo{person}{Alexander Belyaev}, {and}
  \bibinfo{person}{Hans-Peter Seidel}.} \bibinfo{year}{2009}\natexlab{}.
\newblock \showarticletitle{Feature Sensitive Bas Relief Generation}. In
  \bibinfo{booktitle}{{\em IEEE International Conference on Shape Modeling and
  Applications Proceedings}}. \bibinfo{publisher}{IEEE Computer Society Press},
  \bibinfo{address}{Beijing, China}, \bibinfo{pages}{148--154}.
\newblock
\showISBNx{978-1-4244-4068-9}


\bibitem[\protect\citeauthoryear{Li, Wang, Yu, and Ma}{Li
  et~al\mbox{.}}{2012}]%
        {LWYM12}
\bibfield{author}{\bibinfo{person}{Zhuwen Li}, \bibinfo{person}{Song Wang},
  \bibinfo{person}{Jinhui Yu}, {and} \bibinfo{person}{Kwan-Liu Ma}.}
  \bibinfo{year}{2012}\natexlab{}.
\newblock \showarticletitle{Restoration of Brick and Stone Relief from Single
  Rubbing Images}.
\newblock \bibinfo{journal}{{\em IEEE Transactions on Visualization and
  Computer Graphics\/}} \bibinfo{volume}{18}, \bibinfo{number}{2}
  (\bibinfo{year}{2012}), \bibinfo{pages}{177--187}.
\newblock


\bibitem[\protect\citeauthoryear{Sch\"uller, Panozzo, and
  Sorkine-Hornung}{Sch\"uller et~al\mbox{.}}{2014}]%
        {SPS14}
\bibfield{author}{\bibinfo{person}{Christian Sch\"uller},
  \bibinfo{person}{Daniele Panozzo}, {and} \bibinfo{person}{Olga
  Sorkine-Hornung}.} \bibinfo{year}{2014}\natexlab{}.
\newblock \showarticletitle{Appearance-Mimicking Surfaces}.
\newblock \bibinfo{journal}{{\em ACM Transactions on Graphics (proceedings of
  ACM SIGGRAPH ASIA)\/}} \bibinfo{volume}{33}, \bibinfo{number}{6}
  (\bibinfo{year}{2014}), \bibinfo{pages}{216:1--216:10}.
\newblock


\bibitem[\protect\citeauthoryear{Song, Belyaev, and Seidel}{Song
  et~al\mbox{.}}{2007}]%
        {SBS07}
\bibfield{author}{\bibinfo{person}{Wenhao Song}, \bibinfo{person}{Alexander
  Belyaev}, {and} \bibinfo{person}{Hans-Peter Seidel}.}
  \bibinfo{year}{2007}\natexlab{}.
\newblock \showarticletitle{Automatic Generation of Bas-reliefs from 3D
  Shapes}. In \bibinfo{booktitle}{{\em SMI '07: Proceedings of the IEEE
  International Conference on Shape Modeling and Applications}}.
  \bibinfo{pages}{211--214}.
\newblock
\showISBNx{0-7695-2815-5}


\bibitem[\protect\citeauthoryear{Sun, Rosin, Martin, and Langbein}{Sun
  et~al\mbox{.}}{2009}]%
        {SPRF09}
\bibfield{author}{\bibinfo{person}{Xianfang Sun}, \bibinfo{person}{Paul~L.
  Rosin}, \bibinfo{person}{Ralph~R. Martin}, {and} \bibinfo{person}{Frank~C.
  Langbein}.} \bibinfo{year}{2009}\natexlab{}.
\newblock \showarticletitle{Bas-Relief Generation Using Adaptive Histogram
  Equalization}.
\newblock \bibinfo{journal}{{\em IEEE Transactions on Visualization and
  Computer Graphics\/}} \bibinfo{volume}{15}, \bibinfo{number}{4}
  (\bibinfo{year}{2009}), \bibinfo{pages}{642--653}.
\newblock
\showISSN{1077-2626}


\bibitem[\protect\citeauthoryear{S\'{y}kora, Kavan, \v{C}ad\'{i}k,
  Jamri\v{s}ka, Jacobson, Whited, Simmons, and Sorkine-Hornung}{S\'{y}kora
  et~al\mbox{.}}{2014}]%
        {Sykora14}
\bibfield{author}{\bibinfo{person}{Daniel S\'{y}kora},
  \bibinfo{person}{Ladislav Kavan}, \bibinfo{person}{Martin \v{C}ad\'{i}k},
  \bibinfo{person}{Ond\v{r}ej Jamri\v{s}ka}, \bibinfo{person}{Alec Jacobson},
  \bibinfo{person}{Brian Whited}, \bibinfo{person}{Maryann Simmons}, {and}
  \bibinfo{person}{Olga Sorkine-Hornung}.} \bibinfo{year}{2014}\natexlab{}.
\newblock \showarticletitle{Ink-and-Ray: Bas-Relief Meshes for Adding Global
  Illumination Effects to Hand-Drawn Characters}.
\newblock \bibinfo{journal}{{\em ACM Transaction on Graphics\/}}
  \bibinfo{volume}{33}, \bibinfo{number}{2} (\bibinfo{year}{2014}),
  \bibinfo{pages}{16}.
\newblock


\bibitem[\protect\citeauthoryear{Tomasi and Manduchi}{Tomasi and
  Manduchi}{1998}]%
        {TM98}
\bibfield{author}{\bibinfo{person}{C. Tomasi} {and} \bibinfo{person}{R.
  Manduchi}.} \bibinfo{year}{1998}\natexlab{}.
\newblock \showarticletitle{Bilateral Filtering for Gray and Color Images}. In
  \bibinfo{booktitle}{{\em Proceedings of ICCV}}. \bibinfo{pages}{839--846}.
\newblock


\bibitem[\protect\citeauthoryear{Weyrich, Deng, Barnes, Rusinkiewicz, and
  Finkelstein}{Weyrich et~al\mbox{.}}{2007}]%
        {WDBRF07}
\bibfield{author}{\bibinfo{person}{Tim Weyrich}, \bibinfo{person}{Jia Deng},
  \bibinfo{person}{Connelly Barnes}, \bibinfo{person}{Szymon Rusinkiewicz},
  {and} \bibinfo{person}{Adam Finkelstein}.} \bibinfo{year}{2007}\natexlab{}.
\newblock \showarticletitle{Digital bas-relief from 3D scenes}. In
  \bibinfo{booktitle}{{\em SIGGRAPH '07: ACM SIGGRAPH 2007 papers}}.
  \bibinfo{pages}{32}.
\newblock


\bibitem[\protect\citeauthoryear{Wu, Martin, Rosin, Sun, Langbein, Lai,
  Marshall, and Liu}{Wu et~al\mbox{.}}{2013}]%
        {wmr13}
\bibfield{author}{\bibinfo{person}{J. Wu}, \bibinfo{person}{Ralph~R. Martin},
  \bibinfo{person}{Paul~L. Rosin}, \bibinfo{person}{Xianfang Sun},
  \bibinfo{person}{Frank~C. Langbein}, \bibinfo{person}{Y.-K. Lai},
  \bibinfo{person}{A.~David Marshall}, {and} \bibinfo{person}{Y.-H. Liu}.}
  \bibinfo{year}{2013}\natexlab{}.
\newblock \showarticletitle{Making bas-reliefs from photographs of human
  faces}.
\newblock \bibinfo{journal}{{\em Computer-Aided Design\/}}
  \bibinfo{volume}{45}, \bibinfo{number}{3} (\bibinfo{year}{2013}),
  \bibinfo{pages}{671--682}.
\newblock


\bibitem[\protect\citeauthoryear{Xu and Prince}{Xu and Prince}{1998}]%
        {XuP98}
\bibfield{author}{\bibinfo{person}{Chenyang Xu} {and} \bibinfo{person}{Jerry~L.
  Prince}.} \bibinfo{year}{1998}\natexlab{}.
\newblock \showarticletitle{Snakes, shapes, and gradient vector flow}.
\newblock \bibinfo{journal}{{\em IEEE Trans. Image Processing\/}}
  \bibinfo{volume}{7}, \bibinfo{number}{3} (\bibinfo{year}{1998}),
  \bibinfo{pages}{359--369}.
\newblock


\bibitem[\protect\citeauthoryear{Zhang, Zhou, Zhao, and Yu}{Zhang
  et~al\mbox{.}}{2013}]%
        {zzz13}
\bibfield{author}{\bibinfo{person}{Yuwei Zhang}, \bibinfo{person}{Yiqi Zhou},
  \bibinfo{person}{Xiaofeng Zhao}, {and} \bibinfo{person}{Gang Yu}.}
  \bibinfo{year}{2013}\natexlab{}.
\newblock \showarticletitle{Real-time bas-relief generation from a 3D mesh}.
\newblock \bibinfo{journal}{{\em Graphical Models\/}} \bibinfo{volume}{75},
  \bibinfo{number}{1} (\bibinfo{year}{2013}), \bibinfo{pages}{2--9}.
\newblock


\bibitem[\protect\citeauthoryear{Zhang, Chen, Liu, Ji, and Zhang}{Zhang
  et~al\mbox{.}}{2018}]%
        {ZCLJZ18}
\bibfield{author}{\bibinfo{person}{Yu-Wei Zhang}, \bibinfo{person}{Yanzhao
  Chen}, \bibinfo{person}{Hui Liu}, \bibinfo{person}{Zhongping Ji}, {and}
  \bibinfo{person}{Caiming Zhang}.} \bibinfo{year}{2018}\natexlab{}.
\newblock \showarticletitle{Modeling Chinese calligraphy reliefs from one
  image}.
\newblock \bibinfo{journal}{{\em Computers \& Graphics\/}}
  \bibinfo{volume}{70} (\bibinfo{year}{2018}), \bibinfo{pages}{300--306}.
\newblock


\bibitem[\protect\citeauthoryear{Zhang, Zhang, Wang, and Chen}{Zhang
  et~al\mbox{.}}{2016}]%
        {ZZWC2016}
\bibfield{author}{\bibinfo{person}{Yu-Wei Zhang}, \bibinfo{person}{Caiming
  Zhang}, \bibinfo{person}{Wenping Wang}, {and} \bibinfo{person}{Yanzhao
  Chen}.} \bibinfo{year}{2016}\natexlab{}.
\newblock \showarticletitle{Adaptive Bas-relief Generation from 3D Object under
  Illumination}.
\newblock \bibinfo{journal}{{\em Computer Graphics Forum\/}}
  \bibinfo{volume}{35}, \bibinfo{number}{7} (\bibinfo{year}{2016}),
  \bibinfo{pages}{311--321}.
\newblock


\bibitem[\protect\citeauthoryear{Zhang, Zhou, Li, Liu, and Zhang}{Zhang
  et~al\mbox{.}}{2015}]%
        {ZZLLZ15}
\bibfield{author}{\bibinfo{person}{Yu-Wei Zhang}, \bibinfo{person}{Yi-Qi Zhou},
  \bibinfo{person}{Xue-Lin Li}, \bibinfo{person}{Hui Liu}, {and}
  \bibinfo{person}{Li-Li Zhang}.} \bibinfo{year}{2015}\natexlab{}.
\newblock \showarticletitle{Bas-Relief Generation and Shape Editing through
  Gradient-Based Mesh Deformation}.
\newblock \bibinfo{journal}{{\em IEEE Transactions on Visualization and
  Computer Graphics\/}} \bibinfo{volume}{21}, \bibinfo{number}{3}
  (\bibinfo{year}{2015}), \bibinfo{pages}{328--338}.
\newblock


\end{thebibliography}

\end{document}